\documentclass[11pt,a4paper]{article}

\usepackage[utf8]{inputenc}
\usepackage[T1]{fontenc}
\usepackage{mathptmx}          
\usepackage[margin=2.5cm]{geometry}
\usepackage{url}
\usepackage{booktabs}
\usepackage{tabularx}
\usepackage{amsmath}
\usepackage{graphicx}
\usepackage[authoryear]{natbib}
\usepackage{hyperref}
\usepackage{xcolor}
\usepackage{authblk}

\title{VIANA: character Value-enhanced Intensity Assessment\\via domain-informed Neural Architecture}

\author[1,2]{Luana P.\ Queiroz}
\author[2]{Icaro S.\ C.\ Bernardes}
\author[1]{Ana M.\ Ribeiro}
\author[3]{Bernardo M.\ Aguilera-Mercado}
\author[2,4]{Idelfonso B.\ R.\ Nogueira\thanks{Corresponding author}}

\affil[1]{LSRE-LCM, ALiCE, Faculty of Engineering, University of Porto,
  Rua Dr.\ Roberto Frias, 4200-465 Porto, Portugal}
\affil[2]{Department of Chemical Engineering,
  Norwegian University of Science and Technology,
  Sem S\ae lands vei 4, NO-7491 Trondheim, Norway}
\affil[3]{Corporate Fragrance R\&D, The Procter \& Gamble Company,
  Ivorydale Innovation Center, 5299 Spring Grove Ave., Cincinnati, OH 45140, USA}
\affil[4]{Heritage Science Laboratory, Faculty of Chemistry and Chemical Technology,
  University of Ljubljana, Ve\v{c}na pot 113, SI-1000 Ljubljana, Slovenia}

\date{}

\begin{document}
\maketitle

\begin{abstract}
Predicting the perceived intensity of odorants remains a fundamental challenge in sensory science due to the complex, non-linear behavior of their response, as well as the difficulty in correlating molecular structure with human perception. While traditional deep learning models, such as Graph Convolutional Networks (GCNs), excel at capturing molecular topology, they often fail to account for the biological and perceptual context of olfaction. This study introduces VIANA, a novel ``tri-pillar'' framework that integrates structural graph theory, character value embeddings, and phenomenological behavior. This methodology systematically evaluates knowledge transfer across three distinct domains: molecular structure via GCNs, semantic odor character values via Principal Odor Map (POM) embeddings, and biological dose-response logic via Hill's law. We demonstrate that knowledge transfer is not inherently positive; rather, a balance must be maintained in the volume of information provided to the model. While raw semantic data led to ``information overload'' in domain-informed models, applying Principal Component Analysis (PCA) to distill the 95\% most impactful semantic variance yielded a superior ``signal distillation'' effect.
Results indicate that the synthesis of these three knowledge transfer pillars significantly outperforms baseline structural models, with VIANA achieving a peak $R^2$ of 0.996 and a test Mean Squared Error (MSE) of 0.19. In this context, VIANA successfully captures the physical ceiling of saturation, the sensitivity of detection thresholds, and the nuance of odor character value expression, providing a domain grounded simulation of the human olfactory experience. This research provides a robust framework for digital olfaction, effectively bridging the gap between molecular informatics and sensory perception.

\medskip\noindent
\textbf{Keywords:}
Olfactory Intensity Prediction; Domain-Informed Machine Learning; Knowledge Transfer; Sensory Perception
\end{abstract}

\section{Introduction}

The fragrance industry is essential to the global economy, with a market valuation projected to have a revenue of USD 64.47 billion in 2026 \citep{statista_fragrances_2026} and an expected compound annual growth rate driven by a shift toward personalized, sustainable, and high-performance olfactory experiences. Beyond its economic significance, perfumery is a high-stakes engineering discipline where the goal is to design molecular mixtures that elicit precise hedonic and intensity responses. However, olfactory perception remains one of the most challenging modalities to quantify. Unlike vision or audition, which rely on well-defined physical dimensions like wavelength or frequency, olfaction is a multi-modal, non-linear phenomenon. It emerges from the interaction of volatile organic compounds (VOCs) with a complex array of approximately 400 types of olfactory receptors (ORs), making intensity and character difficult to engineer, predict, and standardize across diverse chemical contexts \citep{zarzo2007sense, mainland2014molecule}.

Historically, fragrance development has been a resource-intensive process reliant on the tacit knowledge of skilled perfumers and the logistical burden of extensive sensory panel testing. Although human expertise is fundamental, it faces an insurmountable bottleneck when confronted with the vastness of chemical space. With an estimated $10^{60}$ possible small molecules, only a negligible fraction of which have been characterized for sensory properties. In this context, traditional experimental trial-and-error is no longer a viable strategy for rapid innovation. This discrepancy between the pace of consumer demand and the limits of human evaluation requires the development of robust computational tools for the rapid prioritization and characterization of novel odorants.

Within this context, Machine learning (ML), particularly deep learning on molecular graphs, has emerged as a solution by enabling the mapping of structural topology directly to perceptual outcomes. Leading fragrance houses have begun to integrate AI-driven platforms to accelerate formulation cycles and discover novel ``captive'' molecules \citep{rodrigues2024molecule, santana2023advancing}. Works such as that of \citet{coffin2025machine} presented ML models as a possible solution for interpreting Gas Chromatography-Mass Spectrometry (GC-MS) data to enhance efficiency in ingredient identification, classification, and quality control in the flavor and fragrance industry. This work suggests that conventional chemometric methods require manual data curation, which is time-consuming and error-prone. In view of this, ML could automate complex processes, capture intricate non-linear relationships that traditional statistics miss, and remove the need for peak-by-peak manual spectral matching. Although the suggestions raised in the work by Coffin et al.\ are extremely valuable for improving ML models for the perfume industry, they also raise some troublesome points that warrant consideration, such as data scarcity, heterogeneous or unbalanced datasets that cause overfitting, and strict intellectual property concerns that prevent companies from sharing data to train better models.
In another context of the field, \citet{kim2024nlp} investigated the relationship between descriptive sentences of perfumes and their resulting notes, serving as the core engine for a digital perfume recommendation system. They explore the challenges of previous perfume selection systems that relied on numeric customer satisfaction ratings or outdated text features, applying Advanced Natural Language Processing (NLP) models to deeply understand the semantic nuances of textual marketing descriptions and product reviews, and bridge them with the actual chemical and olfactory notes. Another approach was the one by \citet{fichtelmann2025machine} who claimed that traditional psychophysical models, whether linear or exponential, are limited by scarce and fragmented intensity data, highly variable detection thresholds, and an inability to model receptor saturation. They applied ML models to bypass these limits by capturing high-dimensional structure-perception relationships. The direct prediction approach using an MLP with RDKit descriptors performed best, correctly classifying over 60\% of test instances within their respective categories. The predictions showed a good correlation with actual experimental odor intensity ratings for novel molecules, and they claimed that the models could be improved by integrating other types of information and by explicitly modeling dilution and solvents as covariates. Along this work, we are going to explore this suggestion of integrating other types of information when developing ML models for the perfume industry.

Focusing on predicting properties with more context, we can examine other works in the literature and transpose their suggestions to the perfume industry. For example, \citet{zhao2020prediction}
predicted Critical Heat Flux (CHF) using a hybrid framework that combines domain knowledge (DK) models with machine learning to improve accuracy and generalization. The hybrid approach significantly outperformed standalone ML and DK models, reducing root-mean-square error (RMSE) and showing superior generalization to unseen flow conditions such as high mass flux. In another context, \citet{evans2017predicting} predicted the elastic response, specifically bulk and shear moduli, of pure silica zeolites using geometric descriptors like local geometry, porosity, and structural features alongside a Gradient Boosting Regressor (GBR). This method predicted moduli with accuracy greater than force field approaches and predicted properties for approximately 600,000 hypothetical zeolites.

More closely to the properties that could interest the perfume industry, a notable precedent is the PUFFIN framework by \citet{santana2024puffin}, who introduced the Path-Unifying Feed-Forward Interfaced Network (PUFFIN) framework, which incorporates domain knowledge through an inductive bias node inspired by the Antoine equation to improve vapor pressure prediction. While PUFFIN focuses on physicochemical properties, its innovative use of inductive bias mechanisms has paved the way for advanced sensory modeling, as seen in the ExPUFFIN model presented by \citet{menezes2025expuffin}, which predicted temperature-dependent viscosity of hydrocarbons while enforcing thermodynamic consistency, through the Andrade equation as an inductive bias. This approach reduced RMSE by 37\% compared to a baseline GNN, and it ensured smooth, physically consistent interpolation and extrapolation of viscosity-temperature curves. In a similar vein, the HybridGamma framework by \citet{DICAPRIO2023146104} demonstrated that integrating neural networks with physics-based terms ensures that predicted excess Gibbs energy satisfies the Gibbs-Duhem equation and fundamental boundary conditions; this hybrid approach allows models to leverage data-driven flexibility while strictly adhering to the thermodynamic constraints required to describe complex mixture behavior.

The necessity of this domain-informed strategy becomes even more apparent when confronting the biological complexity of olfactory intensity, which is not encoded by a straightforward rate code of neural spikes \citep{elife, keller2016olfactory}. Instead, total spiking in the olfactory bulb remains normalized across concentrations, while intensity is encoded temporally through the latency and synchrony of neuronal firing in the piriform cortex \citep{elife}. Predicting this response is notoriously difficult for black-box models because concentration-response functions vary significantly between molecules; a compound perceived as dominant at high dilution may become secondary at higher concentrations \citep{koza2005color, martelli2013intensity}. Furthermore, non-monotonic neural responses, multimodal interactions, and high subject variability due to genetic differences in receptors create a sensory landscape that simple data-driven correlations cannot navigate \citep{meredith1986patterned, atanasova2004evaluation}.

Consequently, this article transposes the promising strategy of domain-informed knowledge transfer, proven in the works of PUFFIN, ExPUFFIN, and HybridGamma, directly into the field of perfume engineering. Rather than attempting to predict a single, static point-estimate of intensity, our proposed architecture is trained to predict the parameters of a molecule's entire concentration-response function. By enforcing Hill's law as an architectural inductive bias, we ensure that the model remains anchored to the sigmoidal reality of human olfaction, effectively using established biological logic to compensate for the lack of sensory databases. By adapting these established principles of thermodynamic and physical consistency to sensory data, this work establishes domain information as a critical knowledge pillar for accurate, physically plausible intensity assessment.

Beyond purely physicochemical modeling, recent literature has begun exploring knowledge transfer specifically for perfume-based applications to address the complexity of sensory data. A significant precedent is the work of \citet{oliveira2023framework}, who predicted odor detection thresholds (ODTs) by utilizing a two-stage transfer learning approach. By first training a GCN to recognize semantic odor descriptors, such as ``fruity'' or ``floral'', and subsequently extracting those latent embeddings to initialize a secondary network for threshold prediction, they validated the use of odor character as a critical intermediate ``knowledge bridge'' to reduce prediction errors. Similarly, the scope of transfer learning has extended to molecular generation and substitute discovery. \citet{rodrigues2024molecule} utilized Gated Graph Neural Networks (GGNN) to generate fragrant molecules with targeted notes and physical properties, such as vapor pressure, by iteratively retraining base models on small, note-specific subsets.

While the industry is beginning to harness knowledge transfer, most current models treat olfactory intensity as a static, descriptor-based property, failing to account for the fundamental dose-response nature of scent. Perceived intensity is not merely a function of structure; it is governed by a molecule's gas-phase concentration, dictated by its vapor pressure, and its sigmoidal saturation kinetics at the receptor level \citep{wakayama2019method, sirotin2015neural}. The literature currently lacks a comprehensive framework that simultaneously accounts for all three dimensions of olfaction. This gap suggests significant potential for integrated knowledge transfer strategies, such as the one proposed in the VIANA framework, to move beyond static property prediction and achieve a more holistic characterization of fragrance within the industry.

At the physiological level, the encoding of intensity is inextricably linked to odor character value. Neurobiological evidence suggests that while olfactory receptors recognize multiple odorants through a combinatorial scheme, the brain achieves ``concentration-invariant'' quality recognition by relying on the relative timing and recruitment patterns of glomeruli \citep{zarzo2007sense, mainland2014molecule}. This biological correlation suggests that a model's understanding of ``what'' a molecule smells like (character value) can act as a powerful prior for predicting ``how strongly'' it will be perceived, its intensity. However, existing ML frameworks often treat these tasks in isolation, missing the opportunity for multi-modal knowledge transfer \citep{oliveira2023framework, heid2023chemprop}, which will be explored in this article.

With this, the methodology of this article builds on earlier work, such as that of \citet{wakayama2019method}, which proposed the use of dose--response curve databases for predicting odour intensity. However, it extends this principle into a machine learning framework capable of generalising across molecules and tasks. The objective of this article is to present VIANA (character Value-enhanced Intensity Assessment via domain-informed Neural
Architecture), a domain-informed AI model for olfactory intensity prediction that addresses these limitations through the testing of the combination of three distinct knowledge transfer strategies. At its core, VIANA employs Graph Convolutional Networks (GCNs) to encode the structural topology of molecular graphs, transferring the molecular structure knowledge.
To move beyond black-box structural guessing, we incorporate a second layer of knowledge transfer from phenomenological models. Specifically, we introduce an architectural inductive bias in the form of Hill's law, a sigmoidal function characterizing dose-response behavior, to enforce physicochemical plausibility across a wide concentration range.
The third pillar of the framework integrates semantic odor character value into the model by transferring latent embeddings from a pre-trained ``Principal Odor Map'' (POM) \citep{lee2023principal}. This mimics the neural correlation between an odor's quality and its perceived strength. By refining this high-dimensional semantic data through Principal Component Analysis (PCA) to achieve optimal signal distillation, the VIANA model provides a reliable, high-throughput characterization tool that reflects the complex biological reality of human olfaction. This approach offers unique insights into the advantages and disadvantages of each knowledge transfer strategy and their potential combinations for property prediction in sensorial engineering.

\section{Methodology}

\subsection{Intensity Prediction}
The perceived intensity of an odorant represents a complex, non-linear mapping of physical concentration to a psychophysical percept. It is not an innate or static property of a chemical molecule but rather a dynamic phenomenon emergent from the interaction of volatile organic compounds with approximately 400 types of functional olfactory receptors \citep{zarzo2007sense, mainland2014molecule}. Empirically, sensory strength is characterized as a sigmoidal function of the stimulus logarithm. While linear or exponential functions may approximate this over middle ranges, Hill's Law is currently recognized as the most suitable model for a diverse array of odorants. This equation defines predicted intensity ($\Psi_i$) based on the gas-phase concentration ($X_i$) through three critical parameters: the maximum perceived intensity at saturation ($I_{max}$), the concentration at half-maximal intensity ($C$), and the Hill exponent governing slope and saturation kinetics ($D$), all of which serve as the mathematical foundation for the domain-informed node in the VIANA framework \citep{wakayama2019method}.

However, the underlying neurological encoding is far from a simple rate-based code. Research indicates that total neural spiking in the olfactory bulb often remains normalized across different concentrations \citep{mainland2014molecule}. Instead, the brain utilizes a temporal multiplexing strategy where odor identity is represented by specific ensembles of activated neurons, while intensity is encoded through response latency and population synchrony. As concentration increases, the spatial extent of activated glomeruli expands and response latencies systematically decrease, resulting in more rapid and synchronous neuronal firing \citep{elife}.

This neurological foundation is complicated by a massive psycho-neurological component that drastically alters perceived intensity between individuals. Biologically, humans possess highly varied functional receptor repertoires, leading to unusually large baseline variability in sensitivity. Cognitive and demographic factors, including emotional state, the presence of depressive symptoms, and cultural background, actively modify subjective ratings. Pronounced gender differences also persist; during olfactory training, women consistently demonstrate a more significant increase in self-rated intensity over time compared to men \citep{chao2022gender}. Furthermore, multimodal cognitive cues can ``trick'' the brain, such as when visual color artificially enhances perceived intensity smelled orthonasally but suppresses it retronasally due to an ``intensity contrast'' effect between visually induced expectations and actual physical stimuli \citep{koza2005color}.

Because intensity is profoundly subjective and vulnerable to such biases, quantifying it accurately is a formidable challenge filled with inherent uncertainty. Traditional psychophysical methods, such as matching intensity to standardized chemical scales or using cross-modality techniques like finger-span matching, are plagued by significant within-subject and between-subject variability. Ultimately, verifying absolute truth in these measurements is nearly impossible, as there is no external, objective measurement for the subjective percept \citep{atanasova2004evaluation}. This complexity is exacerbated by a historical paucity of robust empirical data. For decades, computational models relied on isolated datasets, such as the 1985 Dravnieks atlas \citep{dravnieks1992atlas} containing only 138 molecules, severely limiting the structural diversity available for training Machine Learning (ML) algorithms \citep{stopfer2003intensity}. Modern databases still exhibit sparse distribution, as very few molecules are comprehensively evaluated across wide concentration ranges \citep{shang2017machine, keller2016olfactory}. Recently, crowdsourced initiatives like the DREAM Olfaction Prediction Challenge have provided larger datasets (e.g., 476 structurally diverse molecules evaluated by 49 subjects) to improve predictive fidelity \citep{li2018accurate, keller2017predicting}. Using these, researchers have applied algorithms such as Random Forests (RF), Support Vector Machines (SVM), and regularized linear models. RF models have proven particularly effective, as their ensemble approach averages multiple decision trees to mitigate the noise and outliers found in highly variable individual ratings.

Ultimately, despite the performance of these statistical algorithms, a purely data-driven approach remains insufficient for fully solving the olfactory stimulus-percept problem. The inescapable baseline uncertainty of subjective measurement dictates that chemical features alone cannot capture the full reality of human olfaction. To accurately and universally model odor intensity, methodologies must transcend purely statistical correlations and incorporate domain-informed biological rules that account for the non-linear saturation kinetics and temporal neural coding of the brain. By integrating molecular structure, semantic character, and phenomenological behavior, the VIANA framework addresses these limitations to achieve a more robust simulation of the human olfactory experience.

\subsection{Structural Knowledge Transfer via Graph Convolutional Networks}
\begin{figure}[htbp]
    \centering
    \includegraphics[width=\linewidth]{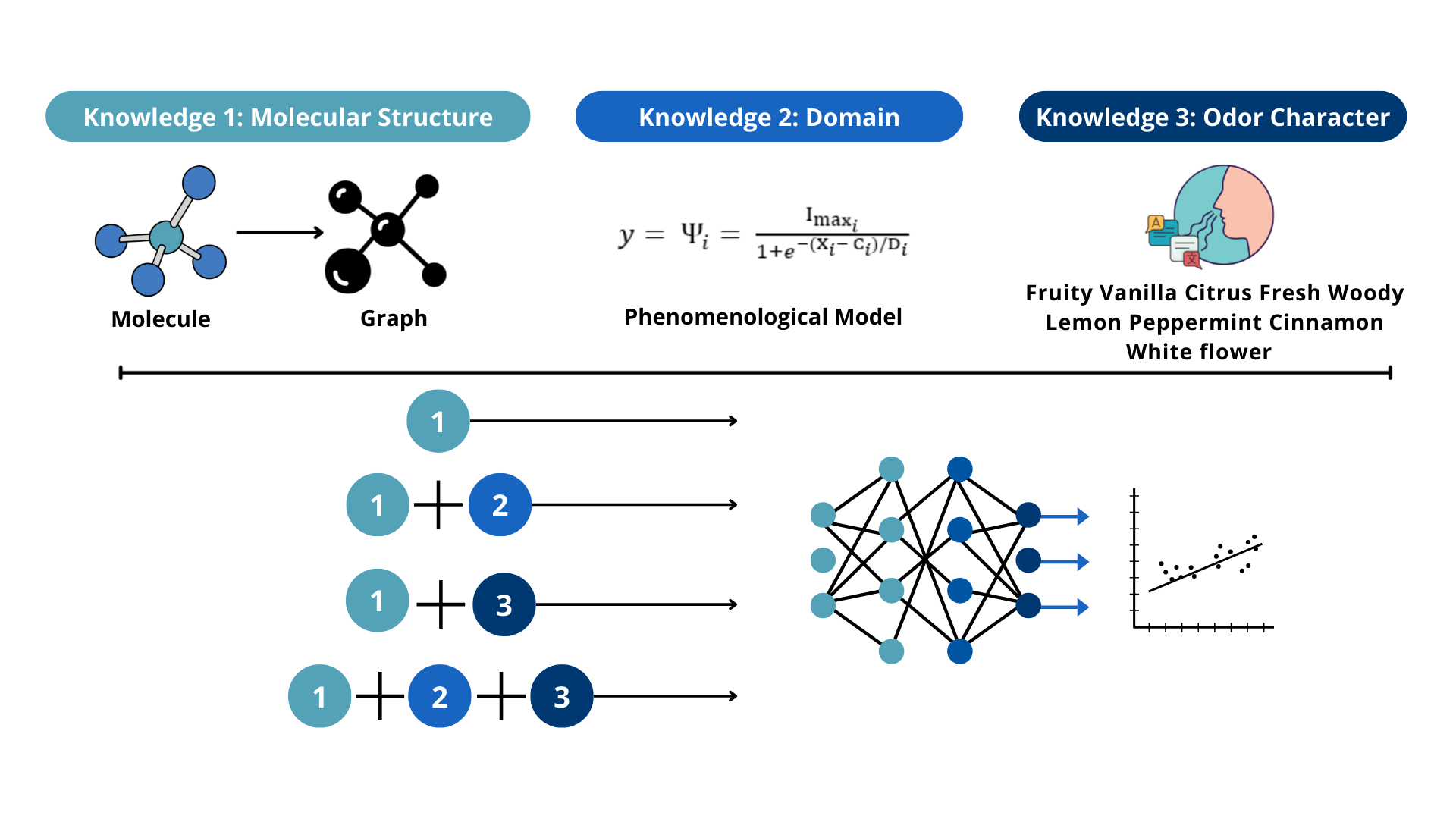}
    \caption{Methodology scheme.}
    \label{fig:Methodology}
\end{figure}

Building upon this understanding of the physiological complexities and data limitations inherent in sensory science, the methodology for this study is structured as a progressive investigation into the impact of different knowledge transfer strategies on the accuracy and physical consistency of odor intensity predictions. As illustrated in Figure~\ref{fig:Methodology}, the framework systematically evaluates the integration of structural, domain-specific, and odor character value knowledge. In this instance, the study evaluates four distinct configurations, including structural knowledge alone, structural combined with domain-informed inductive bias, structural combined with odor character value, and the integration of all three knowledge pillars. Alongside this development, two techniques were implemented: Optuna \citep{optuna_2019} for hyperparameter optimization and model architecture refinement, and Principal Component Analysis (PCA) for dimensionality reduction of character value features. This section describes the progressive pipeline of model construction and the specific mechanisms of knowledge transfer testing.

The first stage of the predictive framework involves encoding the molecular structure into a machine-readable format to test if structural information alone allows a neural network to learn the perceptual patterns established in the database by \citet{wakayama2019method}. Graph Convolutional Networks (GCNs) are utilized to process molecular graphs constructed from SMILES strings, where each molecule is represented as a graph $G = (V, E)$ with $V$ as the set of atoms and $E$ as the set of chemical bonds. GCNs are a specialized class of neural networks designed to process graph-structured data, functioning as a graph-specific variant of Convolutional Neural Networks (CNNs) \citep{zhang2019graph}. Within a GCN, filters are applied to the neighborhoods of each node to extract local features, which are iteratively transformed across layers to form high-level representations. A Laplacian graph matrix typically encodes the topology used in convolution operations to aggregate neighboring information \citep{zhang2019graph, khlifi2023graph}.

The operations within a single GCN layer are defined as follows \citep{khlifi2023graph}:

\begin{enumerate}
\item Calculate the normalized adjacency matrix:
\begin{equation}
\hat{A} = D^{-1/2} (A + I) D^{-1/2}
\end{equation}
where $A$ is the original adjacency matrix, $I$ is the identity matrix, and $D$ is the diagonal degree matrix.
\item Calculate transformed node features:
\begin{equation}
Z = \hat{A} X W
\end{equation}
 where $X$ is the input node feature matrix and $W$ is the trainable weight matrix.
 \item Apply an activation function $f$ to introduce non linearity:
 \begin{equation} H = f(Z) \end{equation}
\item Aggregate neighboring features for node $i$:
\begin{equation}
h_i = \sum_{j \in N(i)} \frac{H_j}{\sqrt{|N(i)| + 1}}
\end{equation} where $N(i)$ represents the neighbors of node $i$.
\item Apply a final linear transformation with weight matrix $W'$:
\begin{equation}
Y = h \cdot W'
 \end{equation}
The weight matrices $W$ and $W'$ serve distinct roles in the network's information flow. While $W$ (Eq.\ 2) acts as a feature extractor that projects the raw input features $X$ into a latent representation $Z$ before graph aggregation, $W'$ (Eq.\ 5) functions as a task-specific projector. It maps the aggregated neighborhood information $h_i$, which now contains the structural context of the graph, into the final output space $Y$ (e.g., class scores). In essence, $W$ learns how to interpret individual node attributes, whereas $W'$ learns how to interpret the combined influence of a node's social neighborhood.

 \item Optionally, apply a softmax function:
\begin{equation}
P = \mathrm{softmax}(Y)
 \end{equation}
\end{enumerate}

\begin{figure}[htbp]
    \centering
    \includegraphics[width=\linewidth]{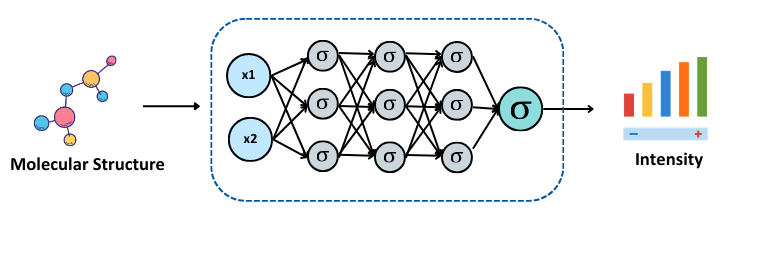}
    \caption{Scheme of the molecular intensity prediction model.}
    \label{fig:GCN}
\end{figure}

In this work, the architecture of the GCN consists of three convolutional layers designed to capture local atomic environments and their global connectivity, implemented through the PyTorch Geometric package \citep{fey_pytorch_geometric}, as illustrated in the transformation from molecular graph to intensity in Figure~\ref{fig:GCN}. The pipeline, then, begins by loading the database containing SMILES strings, the intensity expected values based on \citet{wakayama2019method}, and physicochemical properties such as vapor pressure ($vp$), retrieved via the PubChemPy API \citep{pubchempy_doc}. To address the fundamental dose-response nature of scent intensity, the code generates synthetic data points for each molecule. For molecules with known vapor pressure, a 100-point synthetic dose-response curve is generated, considering the minimum intensity equivalent to 1.3 and the maximum the log of the vapor pressure. Then, the concentration grid ($x$) is established on a log scale, ranging from a calculated detection threshold to the saturation point defined by the molecule's vapor pressure. A scrubbing process removes instances containing missing values in the target intensity ($y$) or input features ($vp$, concentration), resulting in a finalized dataset of 209 molecules.

The dataset is partitioned into training, validation, and testing subsets using an 80, 10, 10 proportion.  SMILES strings are converted into machine-readable graph representations using the \textit{mol2graph} utility, where each node is assigned a 75-dimensional feature vector capturing elemental identity, hybridization, and aromaticity. Molecular topology is preserved via an adjacency matrix and bond attributes. For every point on the generated dose response curve, the structural graph is augmented with two global scalars: the log-transformed vapor pressure and the specific concentration for that data point.

The model is then implemented using a hybrid GCN-MLP architecture. Three GCNConv layers (Graph Convolutional layers) process the molecular graph to learn local atomic environments, after which node-level features are aggregated into a graph-level embedding using a combination of Global Mean Pooling and Global Max Pooling, resulting in a robust structural descriptor. This embedding is concatenated with the standardized vapor pressure and concentration scalars. The resulting fused vector is passed to a Multi-Layer Perceptron (MLP) prediction head utilizing ReLU activation and Dropout to prevent overfitting. The final linear layer includes a clamp operation to ensure predictions remain within physically plausible intensity bounds.

To ensure that the performance differences between knowledge transfer strategies are not artifacts of suboptimal settings, we employ Optuna \citep{optuna_2019}, which is an automated hyperparameter optimization framework designed for machine learning. This framework minimizes the Mean Squared Error (MSE) on the validation set by tuning the learning rate within a range of $1\mathrm{e}{-3}$ to $1\mathrm{e}{-5}$, the dropout rate between 0.1 and 0.4, the number of hidden units in the final fully connected layers chosen from 64, 128, or 256, and the batch size in between 32 or 64. Architecture parity is maintained by keeping the three-layer GCN configuration constant across all comparative tests.

The final predictive model is subsequently trained using the optimal parameter set identified during the 30-trial optimization process. The training architecture utilizes the Adam optimizer integrated with Gradient Clipping, capped at a maximum norm of 1.0, to prevent exploding gradients and ensure numerical stability. To enhance generalization, an early stopping protocol is implemented using a patience-based monitor set to 50 epochs. This mechanism terminates training when the validation loss fails to improve, automatically saving the best-performing model state to disk for final evaluation on the test set. The primary frameworks used throughout the implementation are PyTorch \citep{NEURIPS2019_9015}, PyTorch Geometric \citep{fey_pytorch_geometric}, RDKit \citep{rdkit_cookbook}, and Optuna \citep{optuna_2019}.

Moving beyond a purely structural approach, phenomenological knowledge is introduced through an architectural inductive bias inspired by the physics of olfaction. This involves enforcing Hill's law for the intensity at the output layer of the network to ensure that the model reflects the established knowledge that the intensity expression in olfaction follows an S-shape relative to the logarithm of the stimulus, which fits well to a sigmoidal curve \citep{wakayama2019method}. The inductive bias was implemented based on the logistic equation given by:

\begin{equation}
\Psi_i = \frac{I_{max_i}}{1 + e^{-(X_i - C_i)/D_i}}
\end{equation}

where $\Psi_i$ is the predicted odor intensity and $X_i$ is the concentration of the $i$th odorant (molecule) in the gas phase ($\mu$g/L air). Further, $I_{max_i}$, $C_i$, and $D_i$ are parameters specific to each molecule: $I_{max_i}$ is the odor intensity at saturated gas concentration, $C_i$ is the value of $X_i$ when $\Psi_i$ is $I_{max_i}/2$, and $D_i$ represents the steepness of the function. This architectural choice provides physically consistent predictions across a wide concentration range.

\begin{figure}[htbp]
    \centering
    \includegraphics[width=\linewidth]{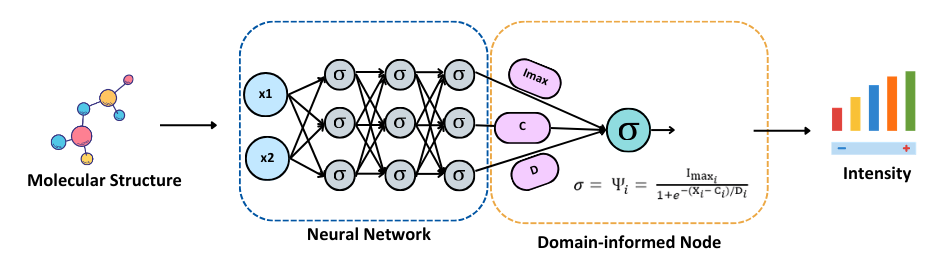}
    \caption{Scheme of the domain-informed molecular intensity prediction model.}
    \label{fig:Domain}
\end{figure}

The primary distinction of this model is the transition from a ``black-box'' regressor to a domain-informed architecture. Instead of allowing the neural network to learn an arbitrary mathematical relationship between molecular structure and intensity, the system enforces a sigmoidal dose-response relationship. To achieve this, significant architectural changes were necessary. In the domain-informed model, the final layer does not output intensity directly; instead, the Multi-Layer Perceptron (MLP) predicts three specific parameters for each molecule, as shown in Figure~\ref{fig:Domain}:

\begin{enumerate}
    \item $I_{max}$ (Asymptote): The maximum possible perceived intensity at saturation.
    \item $C$ (Threshold/Midpoint): The concentration at which half-maximal intensity is reached.
    \item $D$ (Steepness): The Hill coefficient governing the slope of the curve.
\end{enumerate}

Only then is the final output of the model calculated through a constrained activation function:

\begin{equation} \label{eq:Hill}
y = \frac{I_{max}}{1 + e^{-(x - C)/D}}
\end{equation}
where $x$ is the input concentration.

The implementation utilizes the \texttt{F.softplus} function on $I_{max}$ and $D$ to ensure they remain strictly positive, thereby maintaining physical and biological plausibility.

The architecture is deeper and more robust than the previous ``pure'' Graph Convolutional Network (GCN) model to accommodate the added complexity of predicting these sigmoidal parameters. This model utilizes multiple \texttt{GCNConv} layers paired with \texttt{BatchNorm}. The hidden dimensions were increased up to 256, optimized to extract the finer structural features required to define the specific shape of the dose-response curve. While it retains the dual-pooling strategy (Global Mean and Global Max Pooling) of the ``pure'' model to capture both average and prominent molecular features, this version implements \texttt{LayerNorm} within the MLP head. This addition stabilizes the training of the Hill parameters, which are often sensitive to gradient variations.

Moreover, the hyperparameter search is more rigorous. The framework implements an Optuna \texttt{MedianPruner}; if a trial's validation loss at an early epoch is significantly worse than the median of previous trials, the trial is ``pruned'' or terminated. This allows for a more efficient search over 50 trials. The search space was expanded to tune the learning rate ($10^{-5}$ to $10^{-1}$), the dropout rate, the GCN hidden units, and the batch size.

The architectural differences between the GCN baseline model and the domain-informed one are summarized in Table~\ref{tab:comparison}.

\begin{table*}[htbp]
\centering
\caption{Comparison between the ``pure'' GCN and Domain-Informed architectures.}
\label{tab:comparison}
\begin{tabular}{@{}lcc@{}}
\toprule
\textbf{Feature} & \textbf{GCN Model} & \textbf{Domain-Informed Model} \\
\midrule
\textbf{Output Type} & Direct Intensity Prediction ($y$) & Parameters ($I_{max}, C, D$) \\
\textbf{Inductive Bias} & None & Hill's law\\
\textbf{Architecture} & Simple GCN + MLP & GCN + LayerNorm + Sigmoidal Head \\
\textbf{Optimization} & Basic Optuna Search & Optuna with Median Pruning \\
\textbf{Physical Consistency} & Relies on Data & Hard-coded by Architecture \\
\textbf{Stability} & Standard Gradients & Gradient Clipping + LayerNorm \\
\bottomrule
\end{tabular}
\end{table*}

Building upon the initial testing of knowledge transfer from molecular structure and phenomenological domain information, the predictive models are further augmented by odor character value knowledge through the integration of latent embeddings from a pre-trained Primary Odor Map (POM) \citep{lee2023principal}. The POM model is a data-driven Message Passing Neural Network (MPNN), a specialized Graph Neural Network (GNN), designed to map the chemical structure of molecules to their corresponding odor character value. By representing molecules as graphs where atomic and bond properties are explicitly defined, the model dynamically optimizes fragment weights specifically for olfactory applications, thereby surpassing traditional structural fingerprinting techniques.

The POM itself is the high-dimensional representation generated by the network's penultimate layer, which captures a generalized map of structure-odor relationships, while the final layer predicts specific semantic odor qualities, such as ``green'' or ``floral''. Trained on a database of approximately 5,000 labeled molecules, the POM achieves human-level performance; when prospectively tested on hundreds of novel odorants, its predictions matched the consensus of a trained human panel more closely than the median individual panelist did. Furthermore, the POM successfully resolves known structural discontinuities, such as Sell's triplets, where structurally similar molecules exhibit distinct odor profiles, and its output vectors can be mathematically transformed to predict diverse downstream olfactory tasks, like odor detectability, descriptor applicability, and perceptual similarity.

In this work, the embeddings from the penultimate layer of the POM are retrieved and transferred into both the ``pure'' GCN and the domain-informed intensity models to introduce a semantic layer of odor character knowledge. By integrating these features, the models mimic the biological neural correlation between odor quality (character) and perceived strength (intensity). These embeddings are, then, concatenated with the structural features from the primary models before being passed to the prediction layers. Within this framework, the contribution weight of the POM embeddings is treated as a hyperparameter during the model optimization phase to ensure that the character value information does not dominate the training process at the expense of essential intensity features, such as vapor pressure and concentration. The interaction between these semantic, structural, and physicochemical features is refined to maximize predictive accuracy, with weights established via Optuna hyperparameter optimization within a range of 0.1 to 2.

\begin{figure}[htbp]
    \centering
    \includegraphics[width=\linewidth]{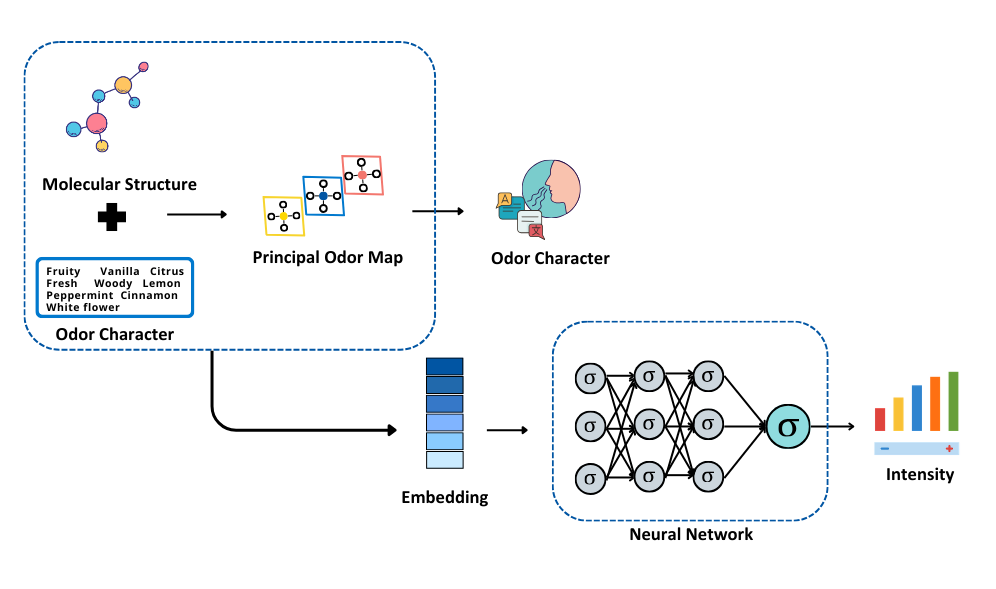}
    \caption{Scheme of the character value-enhanced GCN model.}
    \label{fig:GCNPOM}
\end{figure}

The first implementation of semantic transfer involves augmenting the ``pure'' structural GCN with POM features, as illustrated in Figure~\ref{fig:GCNPOM}. For every molecule in the dataset, a 256-dimensional latent embedding is generated using the pre-trained POM GNN. While the original model relied solely on the GCN structural embedding, this version utilizes a concatenation layer where the 128-dimensional structural vector from the primary GCN is fused with the semantic vector from the POM. The prediction head, consisting of a Multi-Layer Perceptron (MLP), is expanded to process this hybrid vector. Consequently, the first linear layer accepts an input dimension equal to the sum of the structural embedding, the POM components, the log-vapor pressure, and the concentration. Following the logic of its predecessor, this model terminates in a single linear node for direct intensity prediction ($y$), constrained by a ReLU or clamp function to maintain physical plausibility.

\begin{figure}[htbp]
    \centering
    \includegraphics[width=\linewidth]{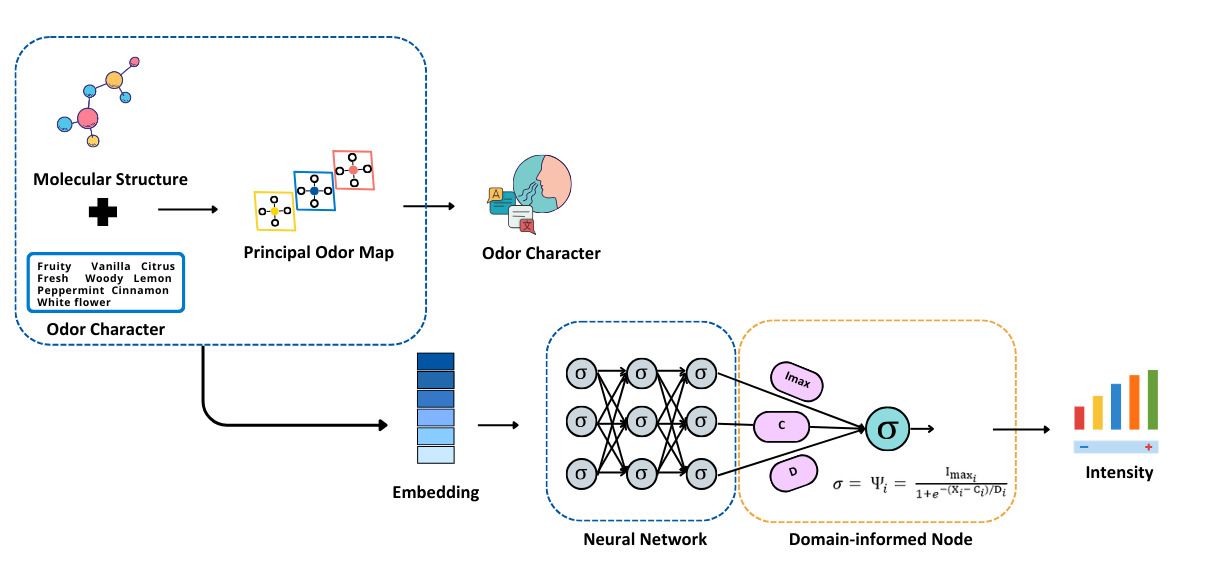}
    \caption{Scheme of the character value-enhanced domain-informed model.}
    \label{fig:DomainPOM}
\end{figure}

The most advanced configuration, shown in Figure~\ref{fig:DomainPOM}, integrates structural, semantic, and phenomenological knowledge into a single cohesive system. This model differs from the original domain-informed model by utilizing the fused structural and POM feature vector to drive the prediction of the Hill's law parameters. In this instance, the MLP head processes the concatenated GCN and POM features to output the maximum intensity ($I_{max}$), the threshold concentration ($C$), and the steepness ($D$). The model continues to employ the \texttt{F.softplus} activation on $I_{max}$ and $D$ to ensure that the predicted sigmoidal curve remains both physically and biologically plausible.

Along this final phase of knowledge transfer implementation, it was necessary to investigate the reduction of dimensionality for the Primary Odor Map (POM) embeddings. While the original POM embeddings provide a rich 256-dimensional representation of odor character, this high dimensionality relative to the limited number of molecules available with experimental intensity data poses a significant risk of overfitting. To mitigate this, Principal Component Analysis (PCA) was utilized to identify the most significant latent dimensions that correlate with sensory intensity. By selecting the top principal components that explain 95\% of the total variance in the semantic odor space, the model retains maximum informativeness while streamlining the feature set.

PCA is a statistical technique designed to reduce dimensionality while retaining the maximum amount of variance, enabling a more manageable representation of a dataset and facilitating a clearer understanding of its underlying structure. Although the method originated before the Second World War, its widespread application in data science began in the 1960s as computational constraints were gradually overcome \citep{pca}. The procedure implemented in this work followed standard statistical steps \citep{pca}:

\begin{enumerate}
\item Standardization of the range of continuous initial variables to ensure each feature contributes equally.
\item Computation of the covariance matrix to unveil linear correlations between variables.
\item Computation of eigenvectors and eigenvalues of the covariance matrix to identify the Principal Components.
\item Creation of a feature vector to determine the most significant principal components.
\item Recasting of the data along the orthogonal axes defined by these principal components.
\end{enumerate}

Mathematically, PCA transforms the dataset into a new coordinate system that maximizes variance along orthogonal directions. In this work, the dimensionality reduction ensures the model remains computationally efficient while retaining core character information.

\begin{figure}[htbp]
    \centering
    \includegraphics[width=\linewidth]{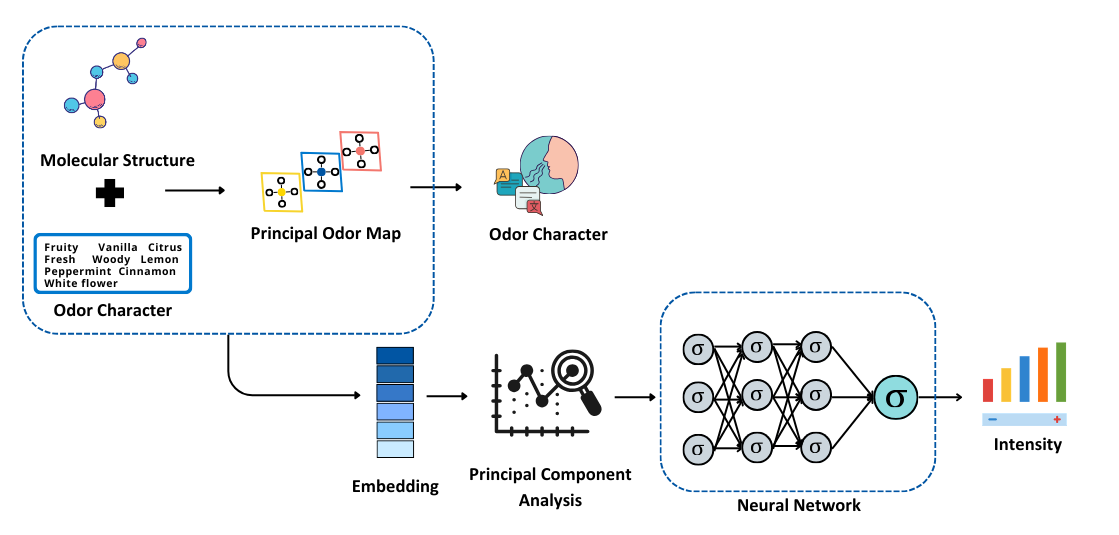}
    \caption{Scheme of the reduced dimensionality character value-enhanced GCN model.}
    \label{fig:GCNPOMPCA}
\end{figure}

In the character value-enhanced structural GCN model, Figure~\ref{fig:GCNPOMPCA}, the input to the prediction head, a Multi-Layer Perceptron (MLP), was adapted to accommodate this reduced feature set. Instead of the original 256-dimensional semantic vector, the first linear layer of the MLP now receives a concatenated vector. This vector consists of the 128-dimensional structural GCN embedding, the specific number of principal components explaining 95\% variance, the log-vapor pressure, and the concentration.

\begin{figure}[htbp]
\centering
    \includegraphics[width=\linewidth]{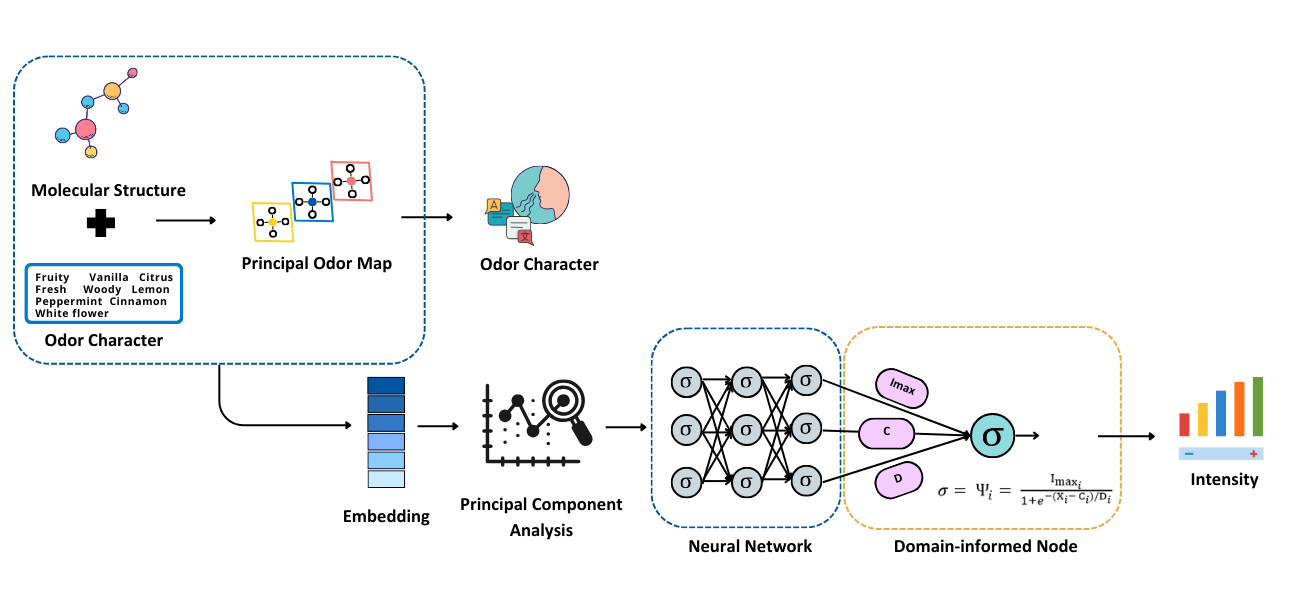}
    \caption{Scheme of the VIANA model.}
    \label{fig:DomainPOMPCA}
\end{figure}

The most complex adaptation occurred in the domain-informed model, Figure~\ref{fig:DomainPOMPCA}, where the fused structural and PCA-reduced semantic features drive the prediction of the Hill's law parameters. The MLP head processes these concatenated features to output $I_{max}$ (maximum intensity), $C$ (threshold concentration), and $D$ (steepness). The use of PCA is particularly vital for this domain-informed approach; the sensitivity of sigmoidal kinetics to input variations necessitates a highly stable and informative feature set. By extracting only the most impactful semantic components, the model achieves better convergence in predicting biological constants that are physically and biologically plausible through the \texttt{F.softplus} activation.

The resulting architecture, which we named VIANA (character Value-enhanced Intensity Assessment via domain-informed Neural Architecture), represents the definitive solution proposed in this work. By fusing topological graph data with semantic odor character and phenomenological dose-response constraints, VIANA provides a biologically grounded, high-throughput tool for the assessment of olfactory intensity across a wide chemical space.

\section{Results and Discussion}

The experimental pipeline was executed within a Google Colab environment utilizing standard CPU resources. As mentioned before, all architectures were implemented using the PyTorch ecosystem, including PyTorch Geometric for graph processing and RDKit for chemical informatics. To ensure rigorous evaluation, the framework systematically investigated the integration of structural, domain-specific, and character value knowledge across six distinct model configurations. The optimization behavior and final performance metrics for these tests are summarized in Table~\ref{tab:results_summary}.

\begin{table*}[htbp]
\centering
\caption{Summary of hyperparameter optimization and training performance across all architectures.}
\label{tab:results_summary}
\resizebox{\textwidth}{!}{%
\begin{tabular}{@{}lcccccccccccc@{}}
\toprule
\textbf{Architecture} & \begin{tabular}[c]{@{}c@{}}\textbf{Opt.}\\ \textbf{MSE}\end{tabular} & \begin{tabular}[c]{@{}c@{}}\textbf{Learning}\\ \textbf{Rate}\end{tabular} & \begin{tabular}[c]{@{}c@{}}\textbf{GCN}\\ \textbf{Hid.}\end{tabular} & \begin{tabular}[c]{@{}c@{}}\textbf{Drop.}\\ \textbf{Rate}\end{tabular} & \begin{tabular}[c]{@{}c@{}}\textbf{Batch}\\ \textbf{Size}\end{tabular} & \begin{tabular}[c]{@{}c@{}}\textbf{POM}\\ \textbf{Weight}\end{tabular} & \begin{tabular}[c]{@{}c@{}}\textbf{Total}\\ \textbf{Epochs}\end{tabular} & \begin{tabular}[c]{@{}c@{}}\textbf{Best}\\ \textbf{Epoch}\end{tabular} & \begin{tabular}[c]{@{}c@{}}\textbf{Val.}\\ \textbf{Loss}\end{tabular} & \begin{tabular}[c]{@{}c@{}}\textbf{Early}\\ \textbf{Stop}\end{tabular} & \begin{tabular}[c]{@{}c@{}}\textbf{Test}\\ \textbf{MSE}\end{tabular} & \textbf{$R^2$} \\ \midrule
GCN (Baseline)  & 33.56   & $9.8\text{e-}4$ & 64  & 0.17 & 32 & N/A  & 120 & 79  & 165.79 & Yes & 195.44 & 0.010 \\
Domain-informed model   & $2.6\text{e-}3$ & $6.8\text{e-}3$ & 64  & 0.24 & 64 & N/A  & 210 & 162 & 0.04   & Yes & 0.46   & 0.991 \\
Character value-enhanced GCN       & 26.89   & $7.6\text{e-}4$ & 256 & 0.12 & 32 & 1.73 & 270 & 220 & 18.67  & Yes & 23.47  & 0.881 \\
Character value-enhanced domain-informed model    & 0.2920  & $1.5\text{e-}3$ & 128 & 0.14 & 32 & 1.26 & 190 & 147 & 0.23   & Yes & 0.55   & 0.989 \\
Reduced dim.\ char.\ value-enhanced GCN & 18.11   & $6.7\text{e-}4$ & 256 & 0.14 & 32 & 1.80 & 470 & 425 & 10.05  & Yes & 33.46  & 0.830 \\
\textbf{VIANA}  & 0.20    & $5.2\text{e-}4$ & 64  & 0.11 & 32 & 0.72 & 170 & 123 & 0.08   & Yes & 0.19   & 0.996 \\ \bottomrule
\end{tabular}%
}
\end{table*}

\begin{figure}[htbp]
\centering
    \includegraphics[width=\linewidth]{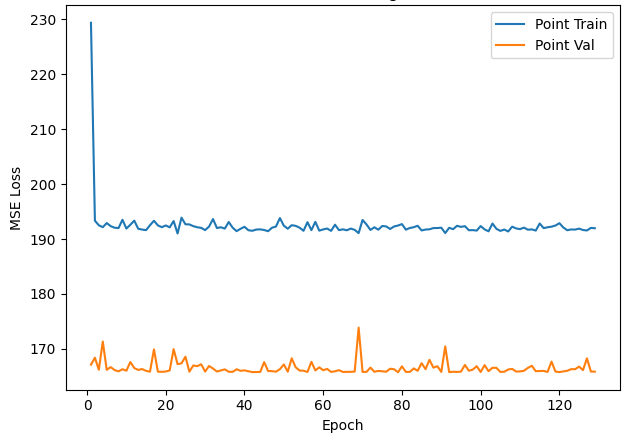}
    \caption{Training and Validation MSE Loss per epoch for the ``pure'' GCN model.}
    \label{fig:GCN_trainval}
\end{figure}

The investigation began with the ``pure'' GCN baseline, which relied strictly on Pillar~1 (Molecular Structure, Figure~\ref{fig:Methodology}). As illustrated in the methodology scheme (Figure~\ref{fig:GCN}), this model lacks any biological or semantic context. This model achieved a final test mean squared error (MSE) of 195.44, representing a poor fit for olfactory intensity. Analyzing the MSE loss per epoch, Figure~\ref{fig:GCN_trainval}, the error stagnated throughout training, showing no meaningful improvement in either training or validation phases.

\begin{figure}[htbp]
\centering
    \includegraphics[width=0.8\linewidth]{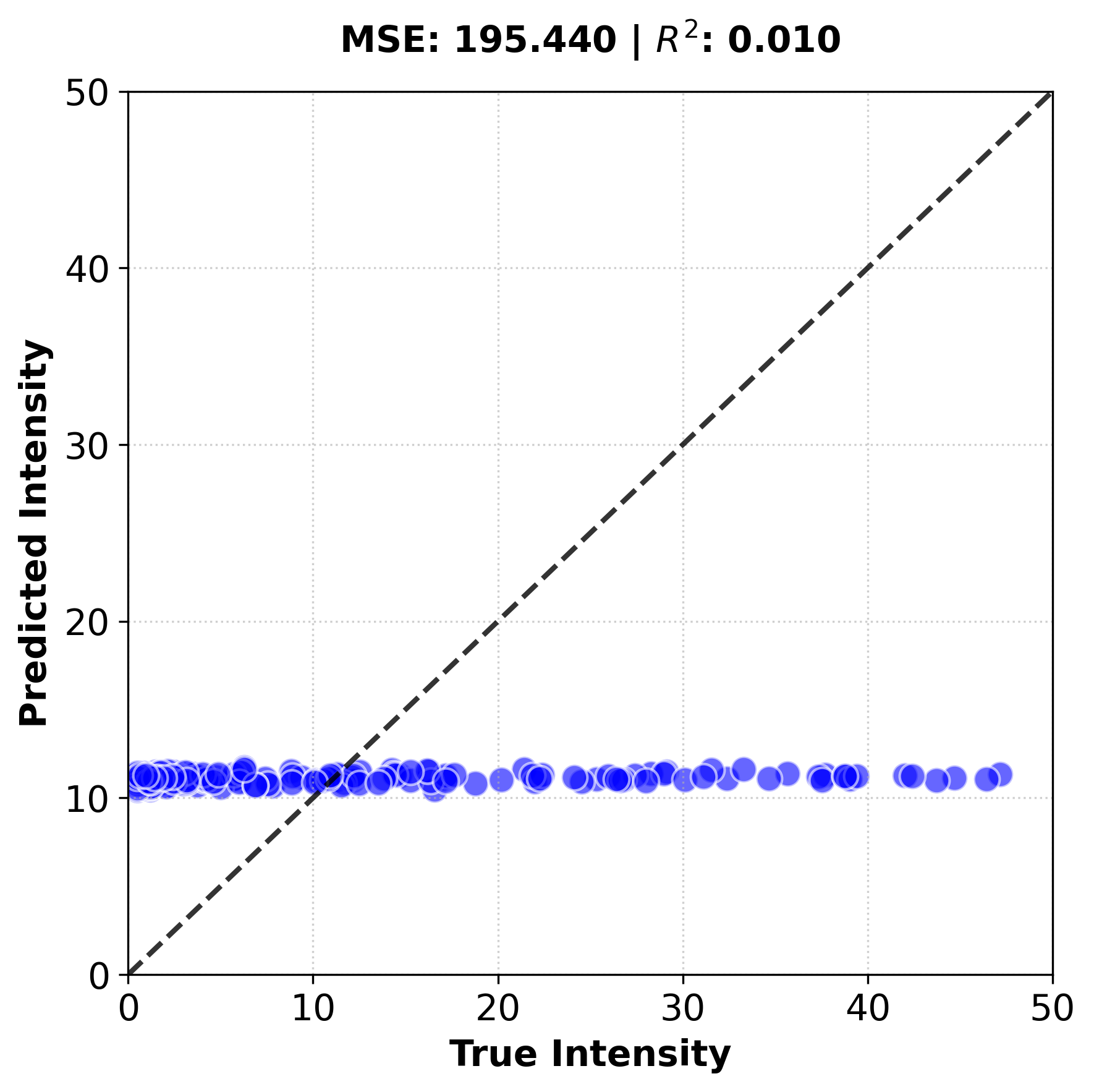}
    \caption{Test set results for the ``pure'' GCN model.}
    \label{fig:GCN_final}
\end{figure}

This failure is visually characterized by a significant ``averaging'' or ``clustering'' effect in the intensity plots, Figure~\ref{fig:GCN_final}; without more context, information, or constraints, the model effectively ``guessed'' intensities near the dataset mean, predicting all points within a narrow range of 10 to 12.

\begin{figure}[htbp]
\centering
    \includegraphics[width=0.8\linewidth]{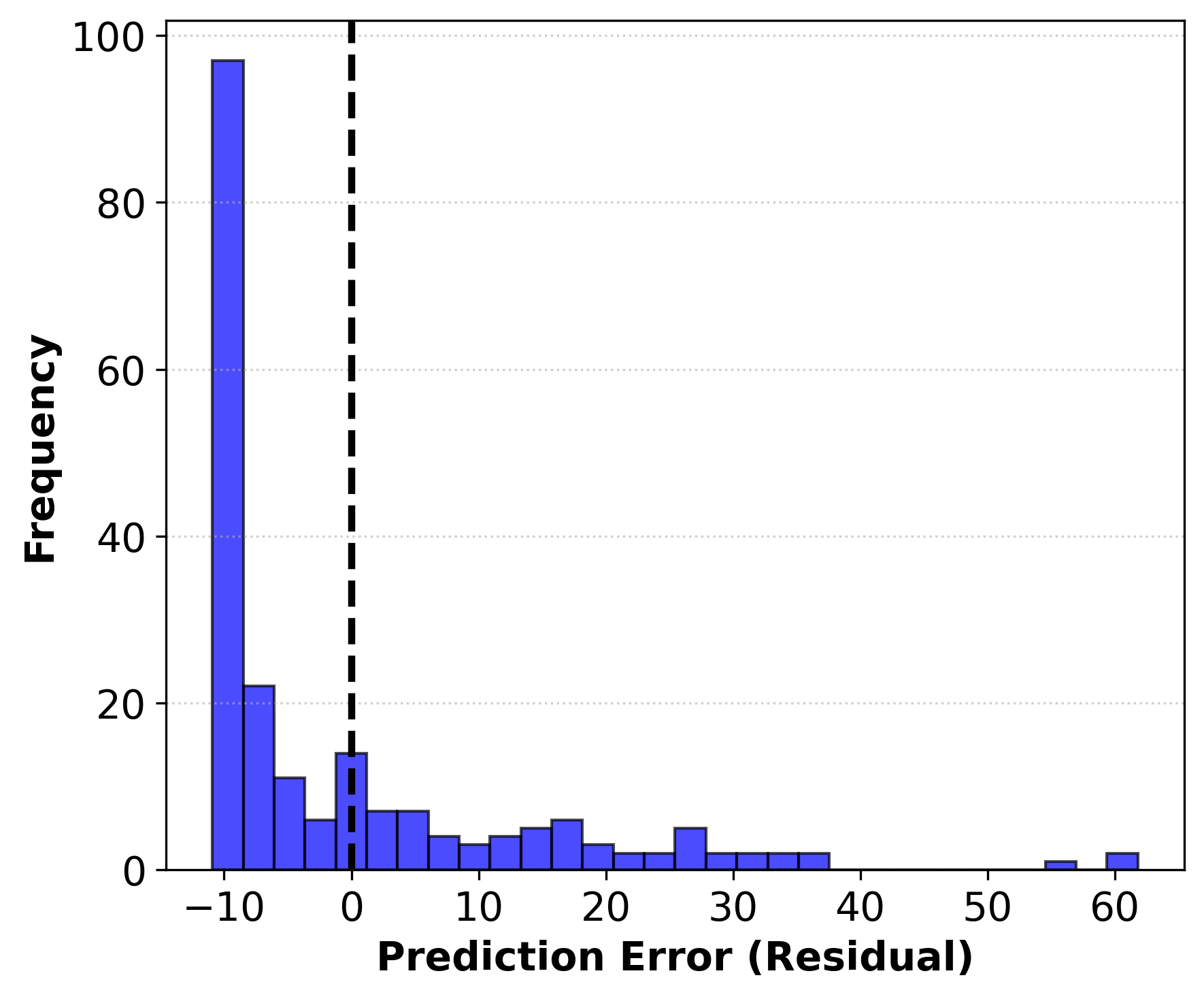}
    \caption{Residual error results for the ``pure'' GCN model training.}
    \label{fig:GCN_error}
\end{figure}

Consequently, the $R^2$ was a negligible 0.010, and the residual error graph, Figure~\ref{fig:GCN_error}, exhibited a wide, scattered dispersion, particularly at the extremes of the intensity scale. These results prove that structural graph theory alone is insufficient to capture the sensory patterns of the dataset, as the model acts as a ``black-box'' regressor that fails to account for biological saturation, leading to physically implausible predictions.

\begin{figure}[htbp]
\centering
    \includegraphics[width=0.8\linewidth]{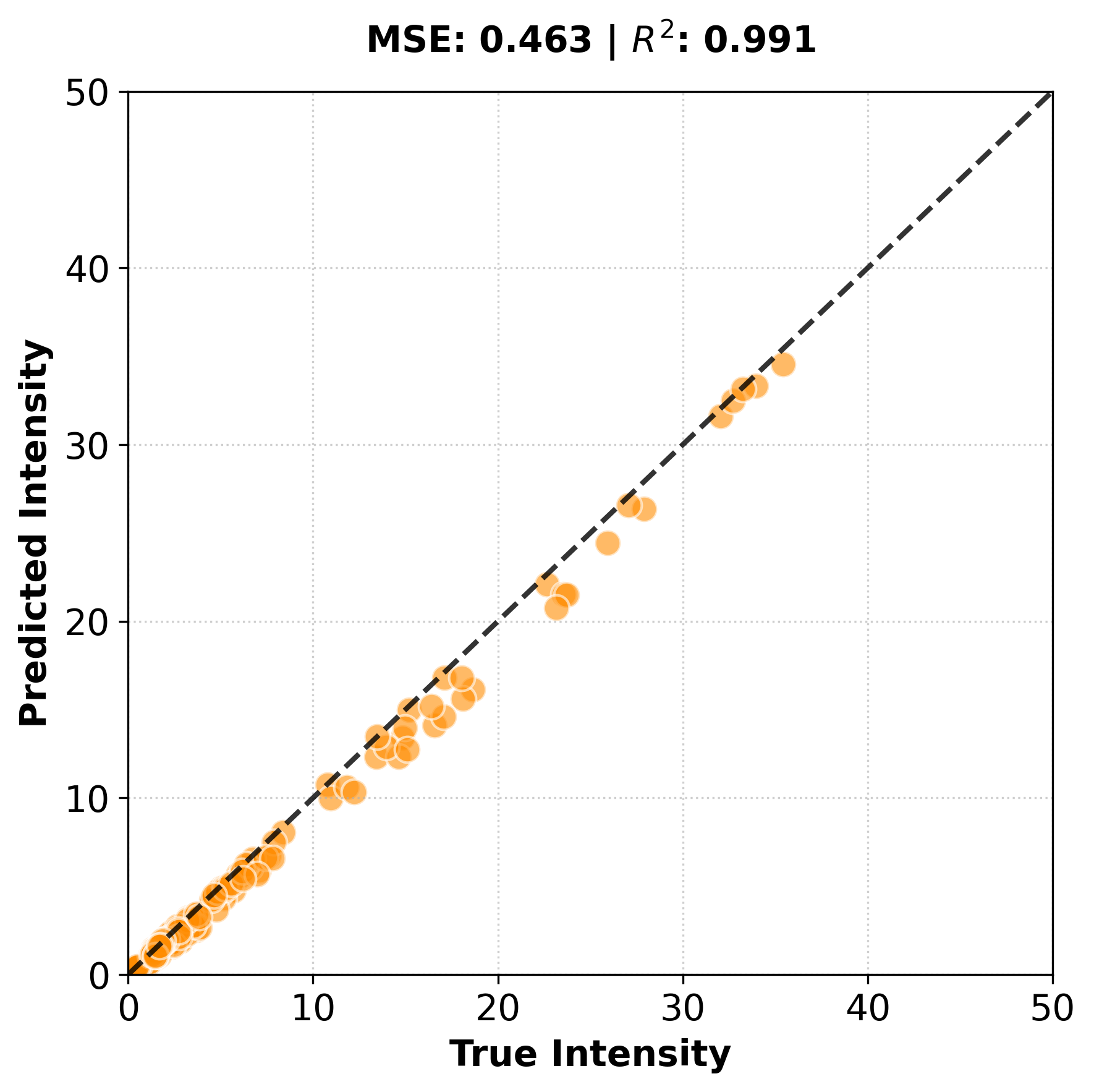}
    \caption{Test set results for the domain-informed model training.}
    \label{fig:Domain_final}
\end{figure}

To address these structural deficiencies, the second stage introduced Pillar~2 (Phenomenological Knowledge, Figure~\ref{fig:Methodology}) via Hill's Law. This configuration follows the domain-informed architecture shown in Figure~\ref{fig:Domain}, shifting the output from raw intensity to biological parameters. This domain-informed model resulted in a dramatic performance gain, reducing the test MSE to 0.46, Figure~\ref{fig:Domain_final}. By enforcing a sigmoidal constraint, the model shifted from predicting raw intensity to mapping chemical features to the phenomenological parameters of perception ($I_{max}, C, D$). This shift acted as a powerful regularizer, ensuring that predictions remained physically possible (e.g., avoiding negative scents or intensities exceeding saturation).

\begin{figure}[htbp]
\centering
    \includegraphics[width=0.8\linewidth]{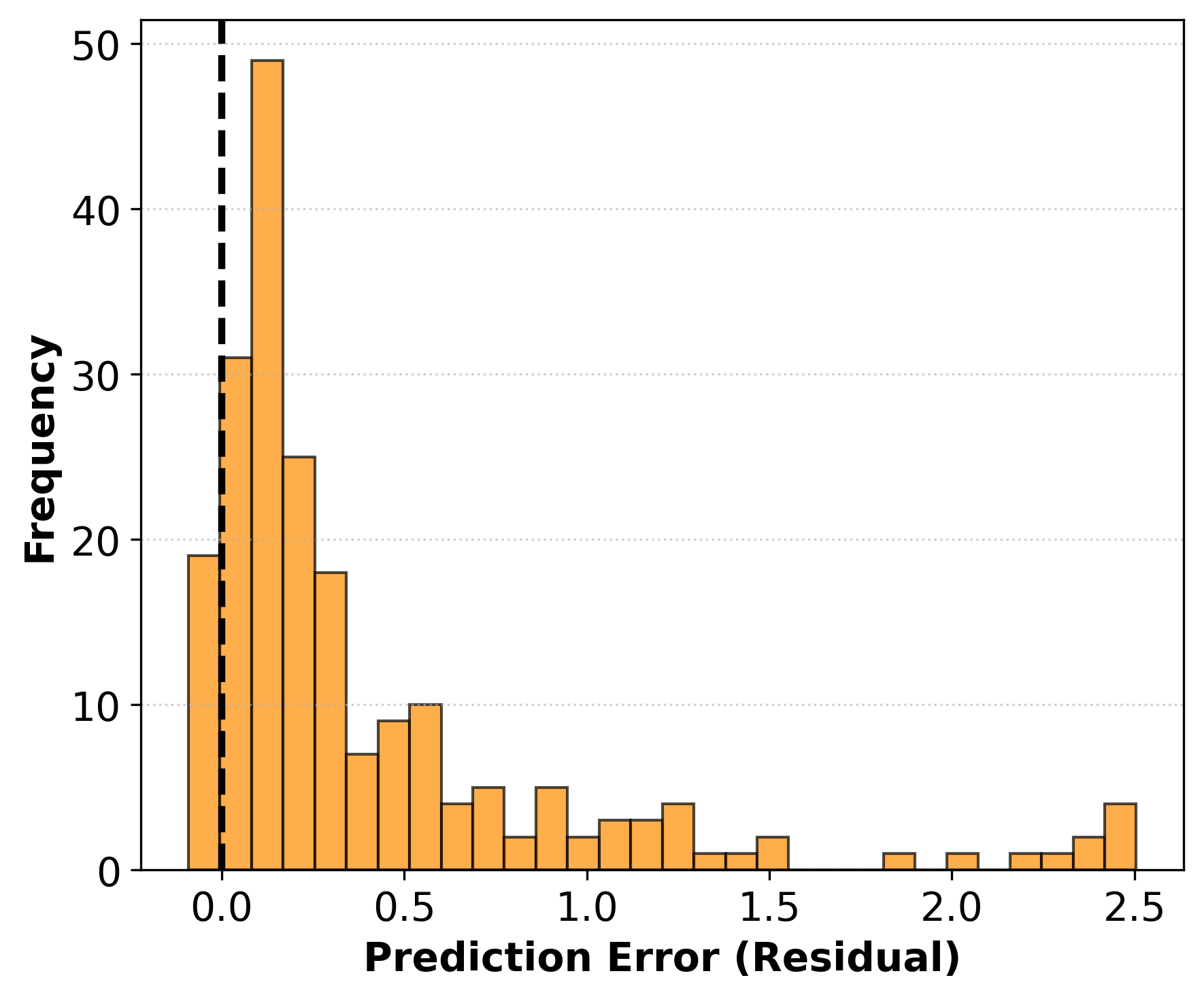}
    \caption{Residual error results for the domain-informed model training.}
    \label{fig:Domain_error}
\end{figure}

The residual error graph, Figure~\ref{fig:Domain_error}, showed a much tighter clustering around the zero axis, and the $R^2$ rose to 0.991, confirming that the sigmoidal bias successfully captured the non-linear variance of the olfactory dose-response behavior.

\begin{figure}[htbp]
\centering
    \includegraphics[width=0.8\linewidth]{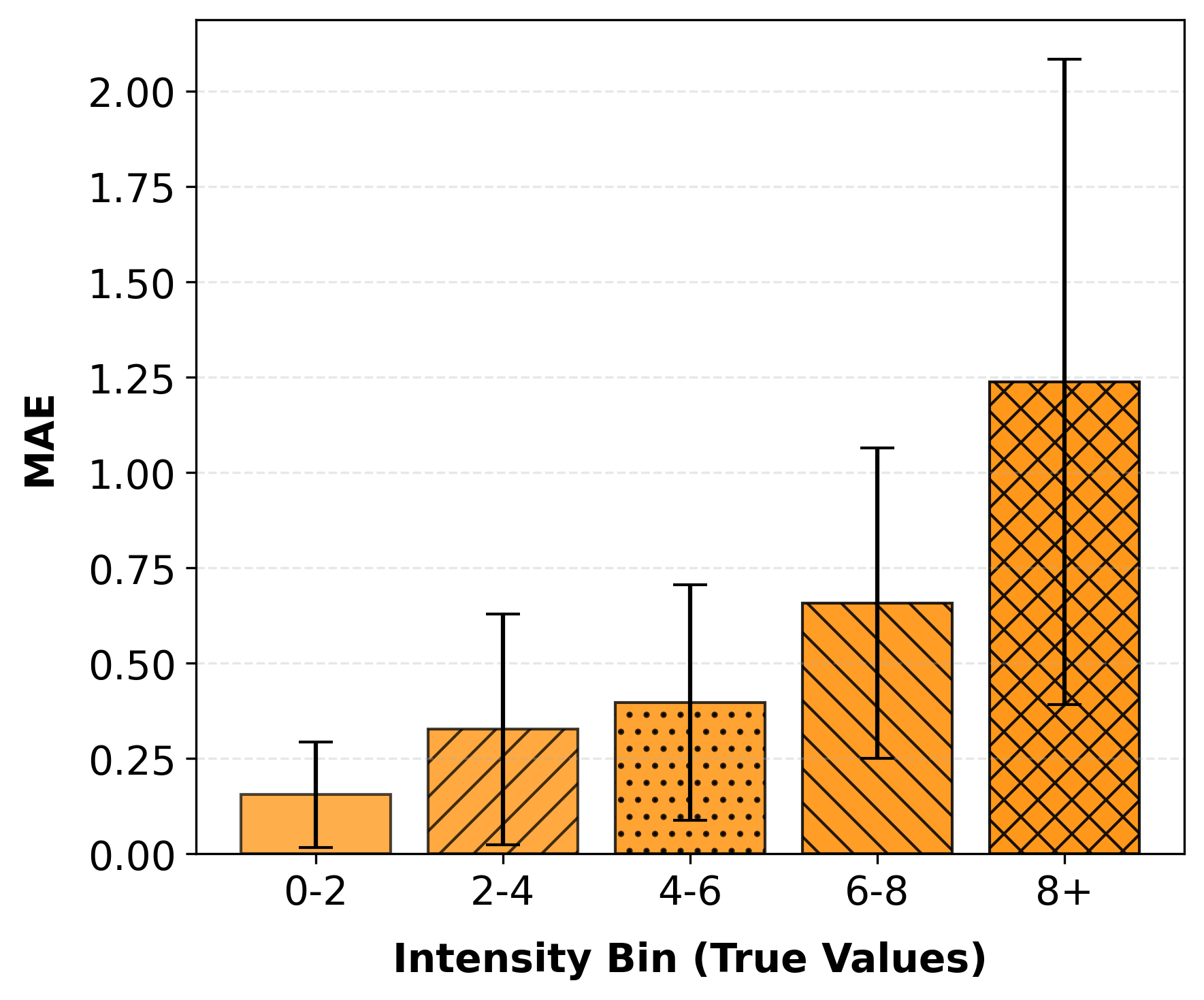}
    \caption{Error distribution per intensity range for the domain-informed model training.}
    \label{fig:Domain_errordistr}
\end{figure}

However, a slight dispersion remained at the highest intensities (8+), suggesting that while the domain bias correctly handled the ``shape'' of the curve, the model still required qualitative ``character'' information to accurately pin down the asymptotic height ($I_{max}$) for complex molecules, Figure~\ref{fig:Domain_errordistr}.

The third and fourth stages of the framework integrated odor character value knowledge via Principal Odor Map (POM) embeddings (Pillar~3, Figure~\ref{fig:Methodology}), as detailed in the architectures of Figure~\ref{fig:GCN} (Structural) and Figure~\ref{fig:Domain} (Domain-informed). This revealed a striking divergence in how structural and domain-informed architectures process qualitative context.

\begin{figure}[htbp]
\centering
    \includegraphics[width=0.8\linewidth]{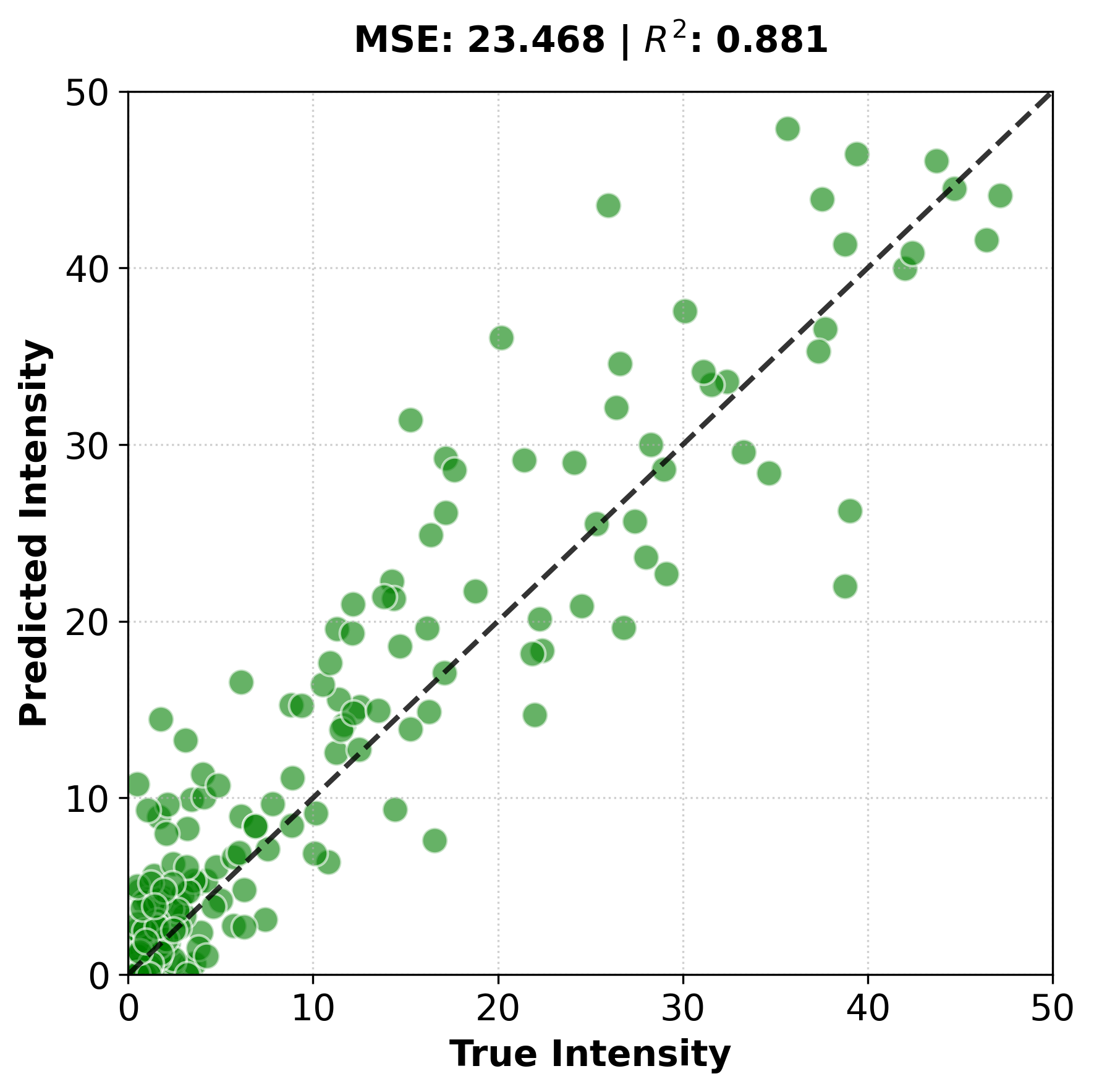}
    \caption{Test set results for the character value-enhanced GCN training.}
    \label{fig:GCNPOM_final}
\end{figure}

For the structural pathway, the integration of POM features into the GCN baseline (character value-enhanced GCN) triggered a major improvement, achieving an MSE of 23.47, a significant reduction from pure structural errors shown in Figure~\ref{fig:GCN_final}, Figure~\ref{fig:GCNPOM_final}.

\begin{figure}[htbp]
\centering
    \includegraphics[width=\linewidth]{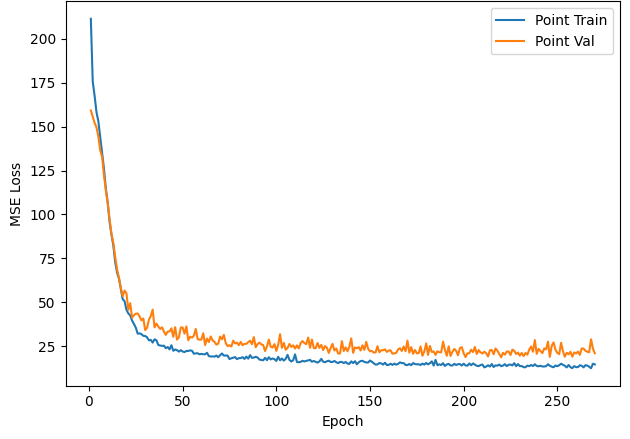}
    \caption{Training and Validation MSE Loss per epoch for the character value-enhanced GCN training.}
    \label{fig:GCNPOM_trainval}
\end{figure}

By equipping the network with an ``odor vocabulary,'' the model moved beyond simple atom-counting to categorize molecules based on character descriptors. For instance, the ability to distinguish a ``musk'' structure from a ``fruity'' one, which correlates strongly with perceived intensity, allowed the model to act as a more informed regressor, Figure~\ref{fig:GCNPOM_trainval}. This is statistically evident in the $R^2$ value, which showed a substantial increase, reflecting the model's newfound ability to explain a much higher percentage of intensity variance.

\begin{figure}[htbp]
\centering
    \includegraphics[width=0.8\linewidth]{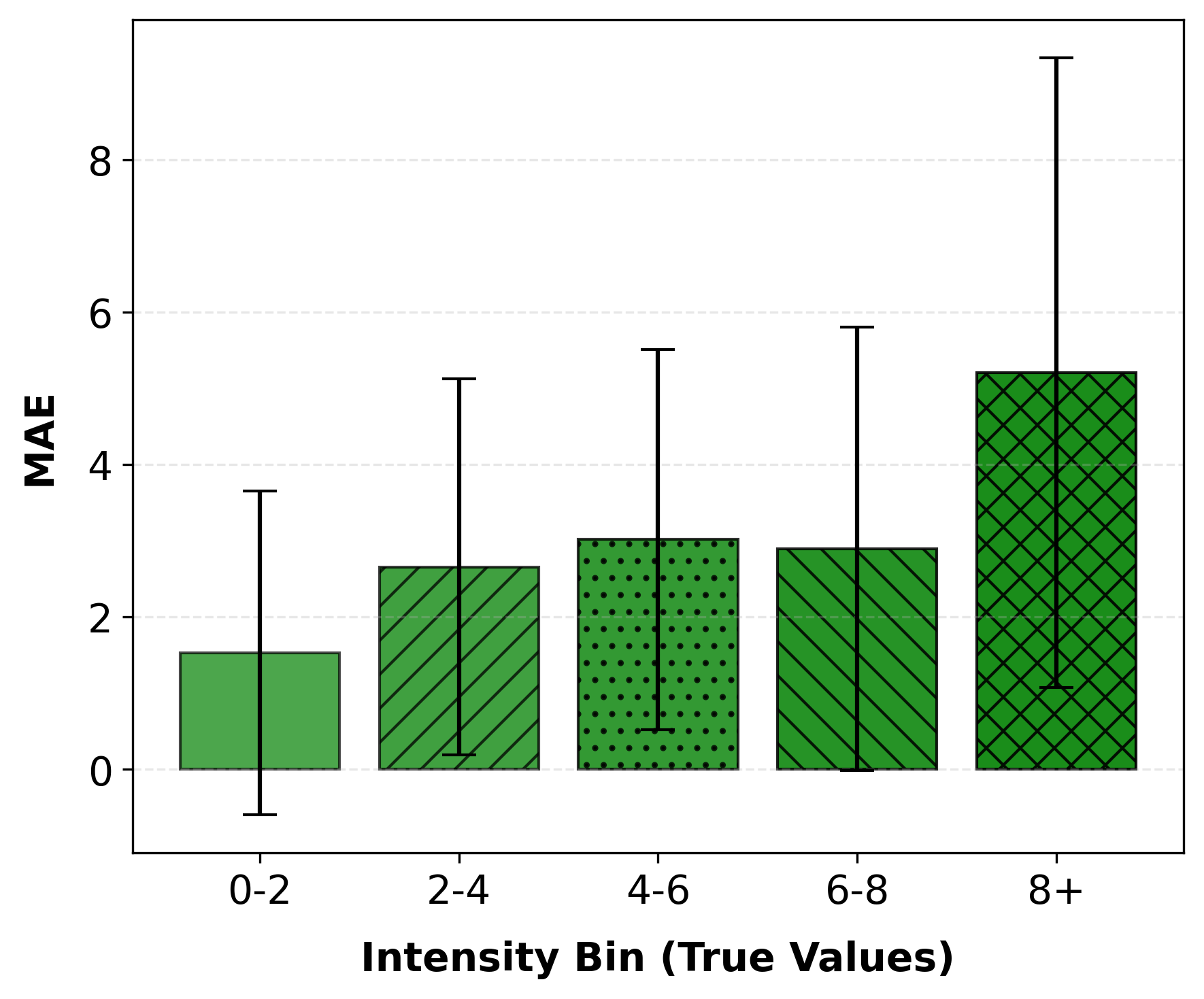}
    \caption{Error distribution per intensity range for the character value-enhanced GCN model training.}
    \label{fig:GCNPOM_errordistr}
\end{figure}

However, diagnostic plots and MAE distributions revealed a critical weakness in this approach, Figure~\ref{fig:GCNPOM_errordistr}. While the Mean Absolute Error (MAE) dropped significantly in the mid-range intensities (4--6), in general the error was dispersed along all the ranges. Without the phenomenological ``ceiling'' provided by Hill's law, the model used the high-dimensional 256-D semantic data to over-extrapolate, leading to physically implausible predictions at high concentrations where the residuals became increasingly erratic.

\begin{figure}[htbp]
\centering
    \includegraphics[width=0.8\linewidth]{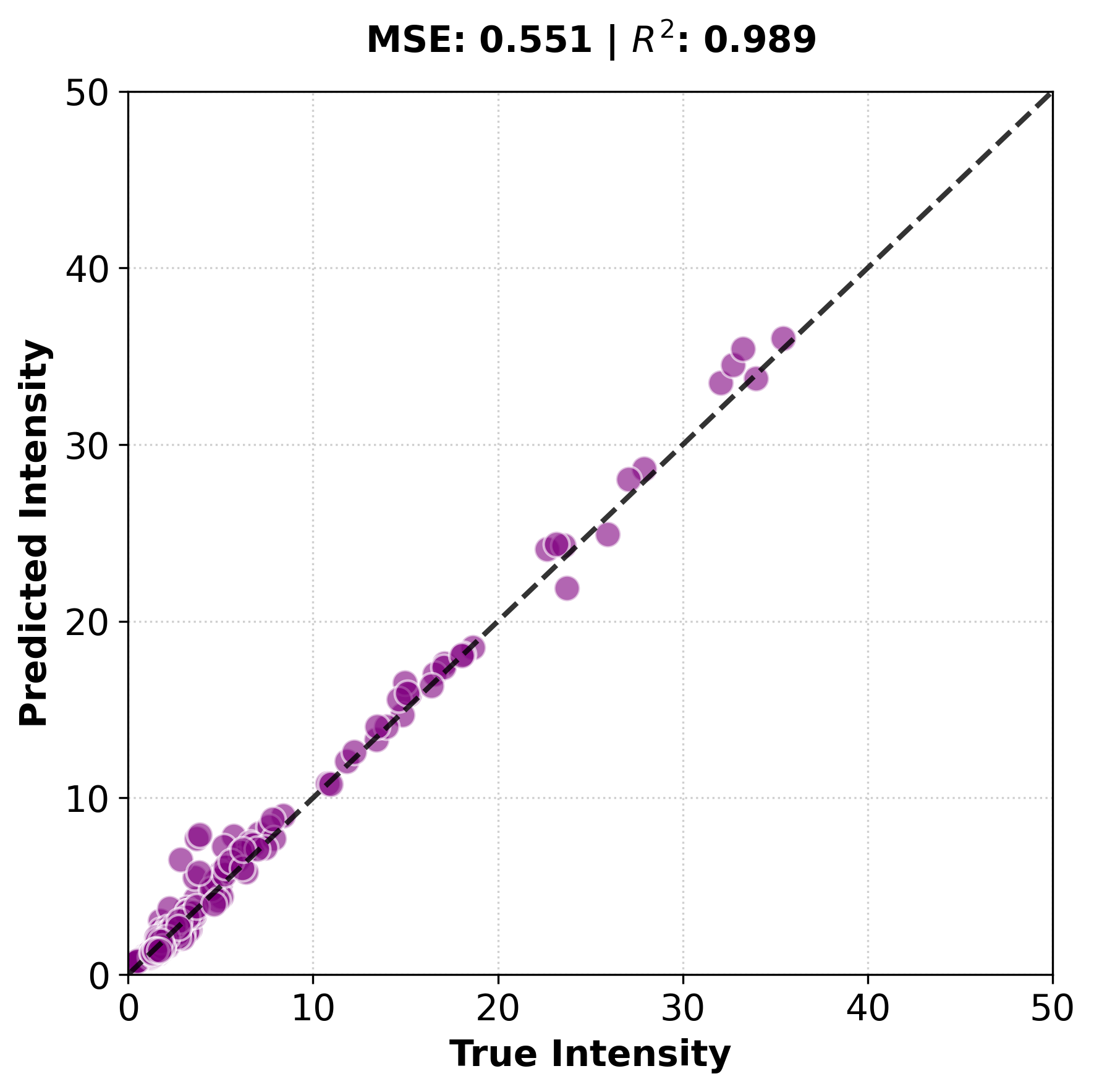}
    \caption{Test set results for the character value-enhanced domain-informed model training.}
    \label{fig:DomainPOM_final}
\end{figure}

In stark contrast, the integration of semantic data into the domain-informed model (Character Value-Enhanced Domain-Informed) initially resulted in a slight performance degradation, with the test MSE rising to approximately 0.55, Figure~\ref{fig:DomainPOM_final}. This phenomenon represents an ``Information Overload'' effect. The Hill's Law parameters ($I_{max}, C, D$) are mathematically sensitive to input variations; the introduction of the raw, 256-dimensional POM embedding created a high-dimensional input space that effectively ``overwhelmed'' the sigmoidal output head.

\begin{figure}[htbp]
\centering
    \includegraphics[width=\linewidth]{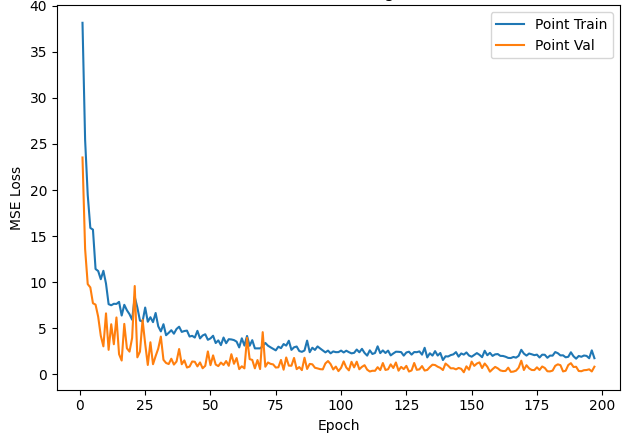}
    \caption{Training and Validation MSE Loss per epoch for the character value-enhanced domain-informed model training.}
    \label{fig:DomainPOM_trainval}
\end{figure}

Analysis of the training and validation loss curves suggests that the abundance of features introduced redundant signals and ``noise'' that competed with critical physical features like vapor pressure, Figure~\ref{fig:DomainPOM_trainval}.

This competition led to less precise parameter estimation and gradient instability during optimization. While the hard-coded inductive bias maintained the characteristic S-shape, the prediction exhibited deviations in plateau height ($I_{max}$) and threshold ($C$) because the MLP was forced to navigate unnecessary semantic complexity. The resulting drop in $R^2$ confirmed that for a domain-informed model, character value features density must be optimized to preserve the mathematical stability of the phenomenological parameters.

\begin{figure}[htbp]
\centering
    \includegraphics[width=0.8\linewidth]{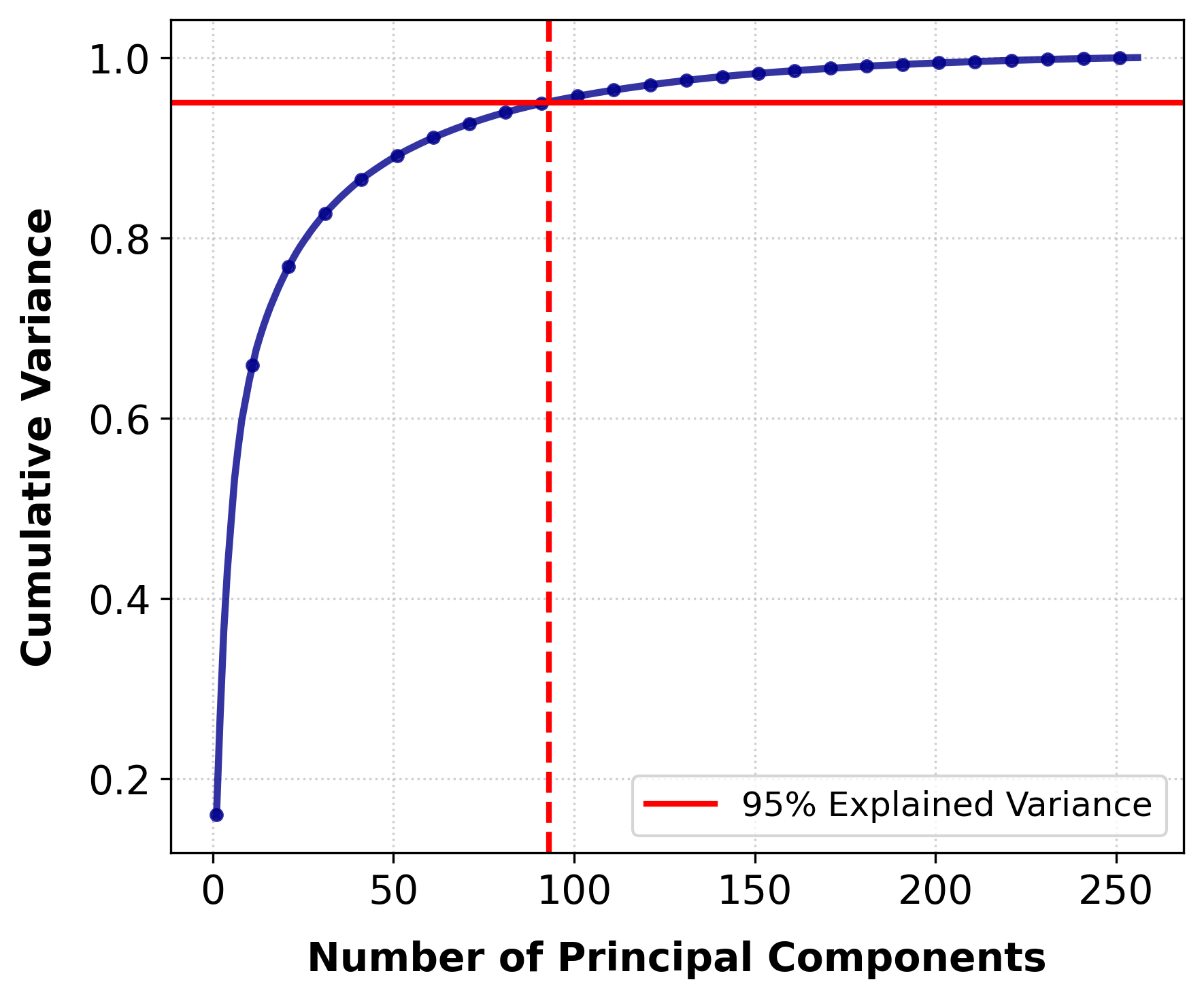}
    \caption{Variance analysis on principal components.}
    \label{fig:PCA}
\end{figure}

The identified limitations of raw character value integration necessitated a final refinement phase: the application of Principal Component Analysis (PCA) to extract 95\% of the semantic variance while discarding latent noise. A variance analysis was conducted to determine the optimal number of components, as illustrated in the scree plot in Figure~\ref{fig:PCA}, identifying that 93 principal components were required to achieve the goal of 95\% cumulative variance.

\begin{figure}[htbp]
\centering
    \includegraphics[width=0.8\linewidth]{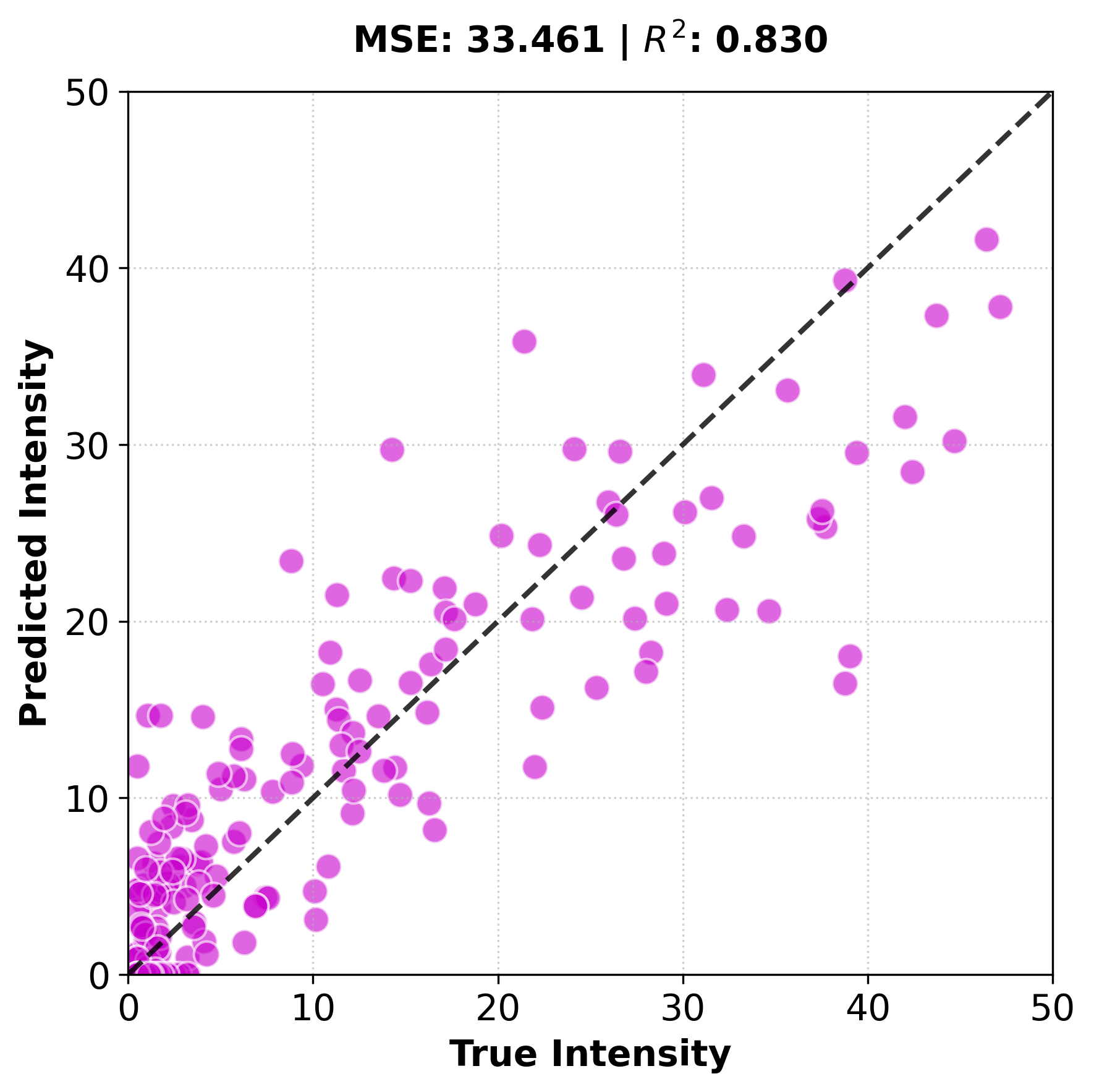}
    \caption{Test set results for the reduced dimensionality character value-enhanced GCN training.}
    \label{fig:GCNPOMPCA_final}
\end{figure}

The impact of this dimensionality reduction was markedly different across the two base architectures. In the GCN-based models, PCA helped densify the input information, facilitating more stable training over an extended 470 epochs. However, this stability came at the cost of predictive performance, Figure~\ref{fig:GCNPOMPCA_final}. For the reduced dimensionality character value-enhanced GCN, the decrease in performance compared to the raw POM version is attributed to the model's reliance on ``brute-force'' feature extraction. Lacking a biological inductive bias (Hill's Law), the ``pure'' GCN requires maximum data granularity to approximate the complex, non-linear relationship between structure and intensity. By compressing the 256-D POM vector, we inadvertently removed ``fine-grained'' semantic details, subtle clues that helped the network map specific chemical families.

\begin{figure}[htbp]
\centering
    \includegraphics[width=0.8\linewidth]{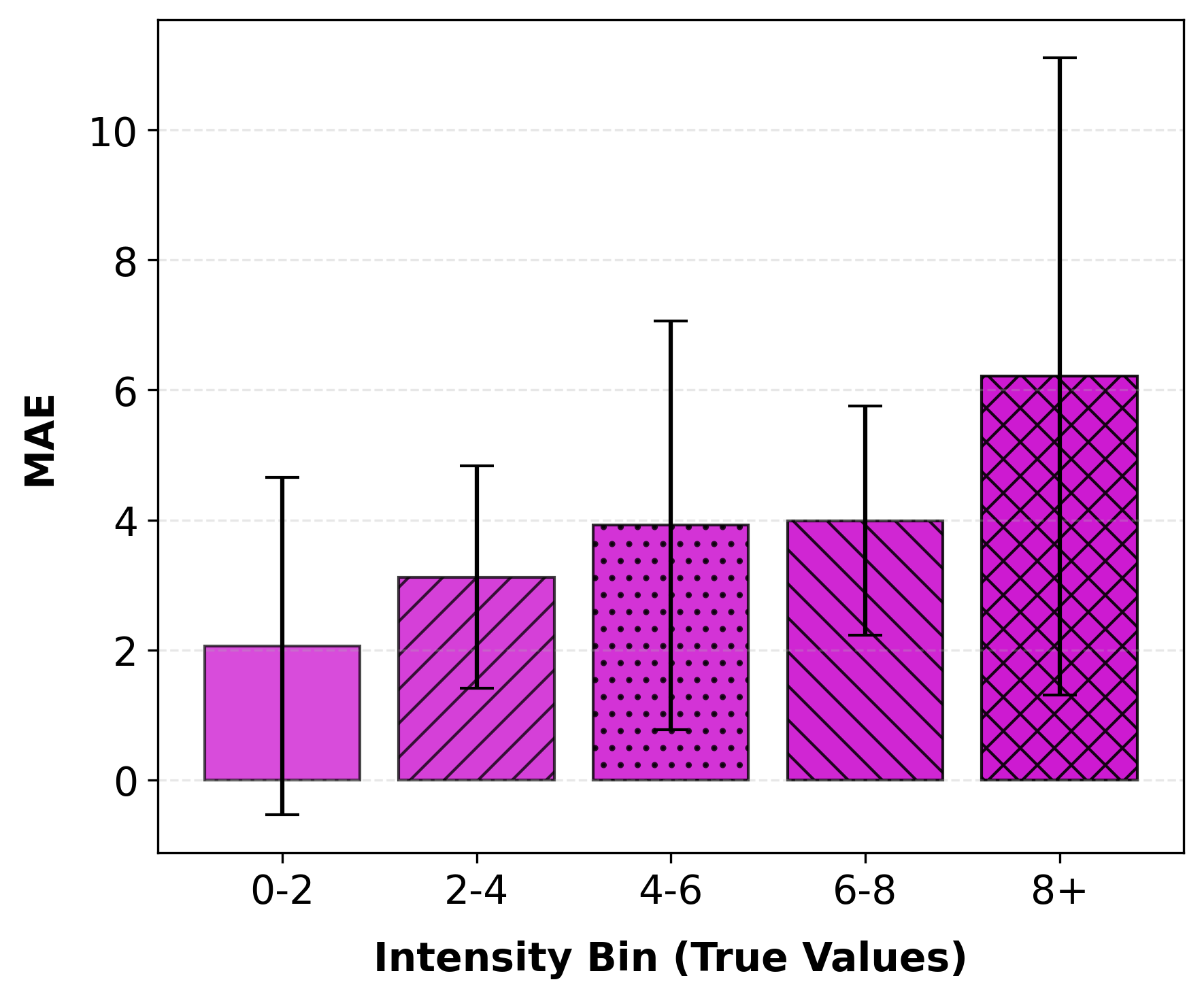}
    \caption{Error distribution per intensity range for the reduced dimensionality character value-enhanced GCN training.}
    \label{fig:GCNPOMPCA_errordistr}
\end{figure}

Consequently, the GCN lost the ``richness'' of the vocabulary it was using to compensate for its lack of physical logic. This is reflected in the increased scatter of the test prediction results and a slight drop in $R^2$. Furthermore, the error distribution, Figure~\ref{fig:GCNPOMPCA_errordistr}, remained high in the extreme bins (0--2 and 8+), as the GCN still lacked the ``physical ceiling'' of saturation, leading to higher errors at the edges of the dataset.

\begin{figure}[htbp]
\centering
    \includegraphics[width=0.8\linewidth]{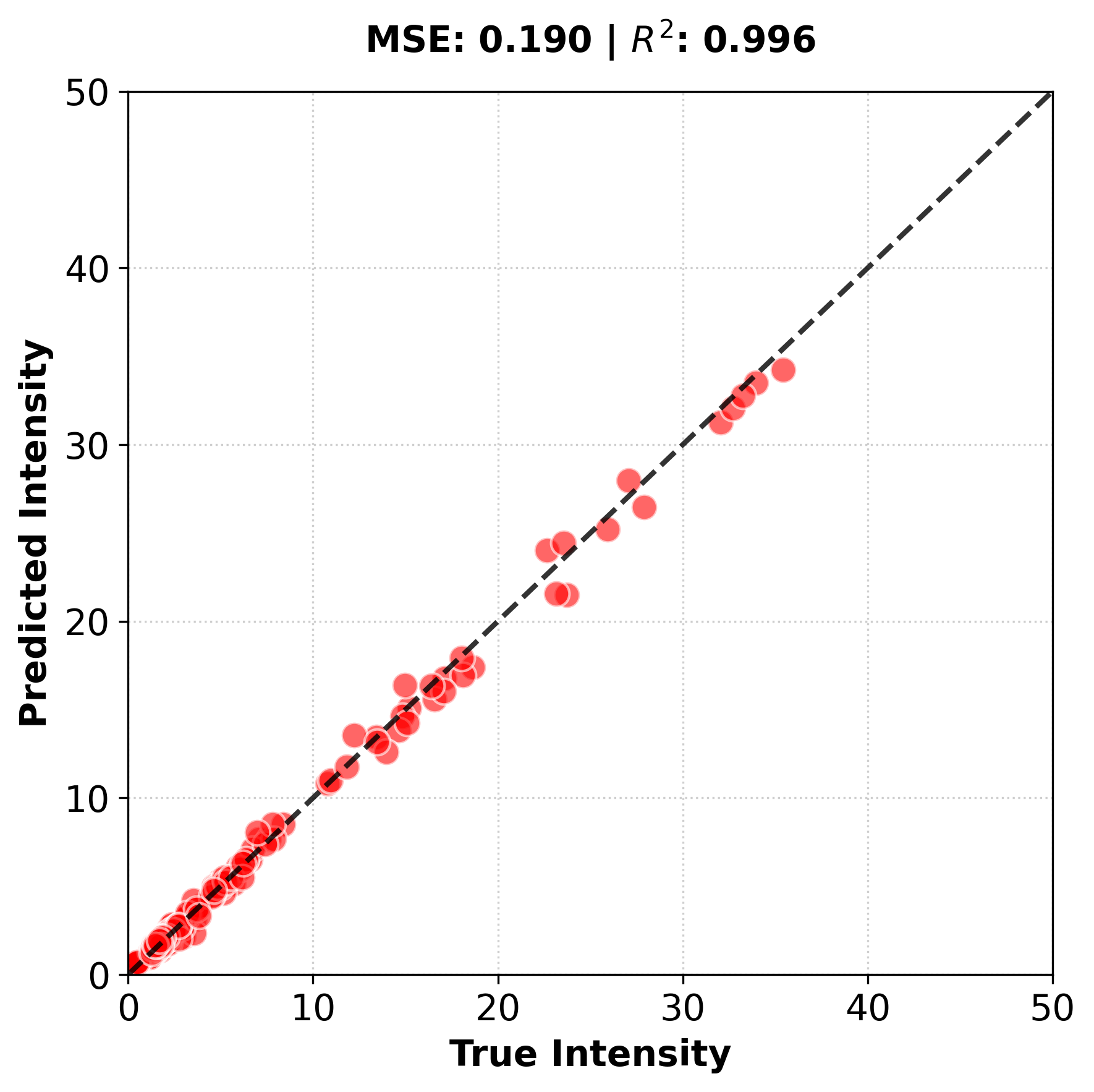}
    \caption{Test set results for the VIANA model training.}
    \label{fig:VIANA_final}
\end{figure}

In contrast, the final solution, the VIANA model (Figure~\ref{fig:DomainPOMPCA}), represents the full integration of all three pillars: molecular structure, odor character, and domain-informed. The VIANA model achieved the best overall performance with a scaled test MSE of 0.19, Figure~\ref{fig:VIANA_final}. By filtering the semantic space through PCA, the model successfully integrated odor character into the biological parameter prediction head without causing the gradient instability seen in previous iterations.

\begin{figure}[htbp]
\centering
    \includegraphics[width=0.8\linewidth]{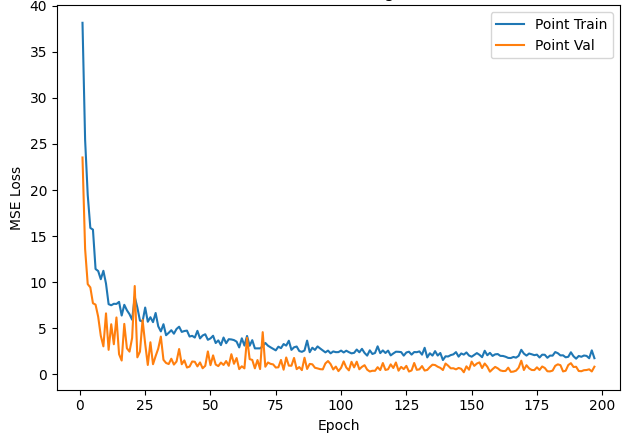}
    \caption{Training and Validation MSE Loss per epoch for the VIANA model training.}
    \label{fig:VIANA_trainval}
\end{figure}

The training and validation loss curves for VIANA exhibit smooth convergence, with early stopping successfully preventing overfitting at epoch 170, Figure~\ref{fig:VIANA_trainval}.

The VIANA model's improvement after PCA represents the definitive outcome of the knowledge transfer testing; this is where the synthesis of knowledge truly pays off. The MLP in the domain-informed model is tasked with predicting three highly sensitive biological constants ($I_{max}, C, D$). When presented with the raw 256-D POM embedding, the network was forced to navigate a ``high-dimensional noise floor'' where redundant and weakly correlated dimensions acted as distractors. PCA acted as a ``distillation'' filter, refining the 256 dimensions into the most impactful semantic signals. For the Hill's law head, this allowed a focus on the ``big picture'' descriptors that dictate the shape of the dose-response curve without the confusion of high-dimensional clutter.

\begin{figure}[htbp]
\centering
    \includegraphics[width=0.8\linewidth]{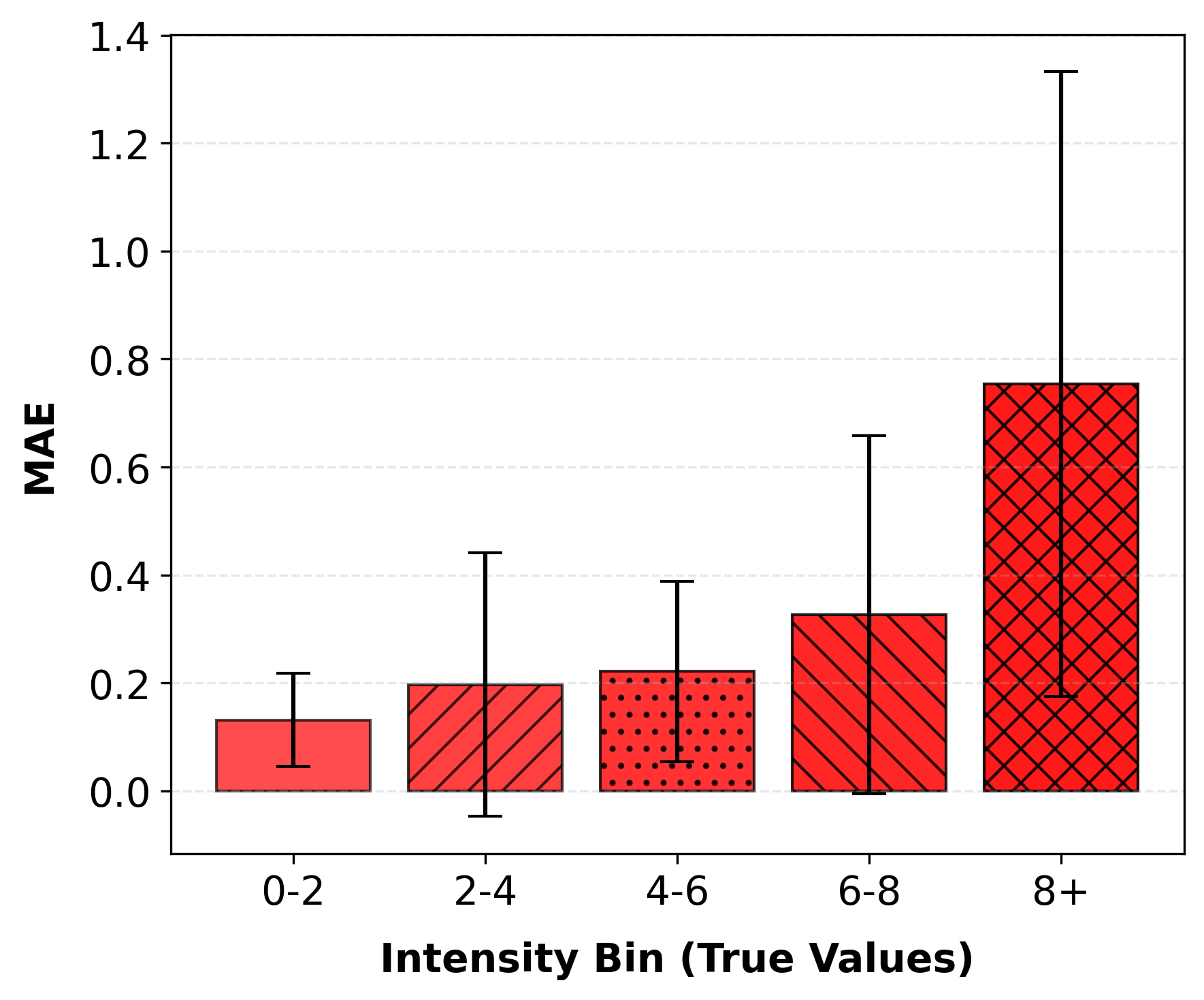}
    \caption{Error distribution per intensity range for the VIANA model training.}
    \label{fig:VIANA_errordistr}
\end{figure}

The results for VIANA are striking: the $R^2$ reached a study-wide peak, confirming that parameters are most accurately predicted when the input ``signal'' is clean. The prediction results show a significantly tighter correlation, and the error distribution is remarkably uniform, Figure~\ref{fig:VIANA_errordistr}. Because the Hill's law hard-codes the ``S-shape,'' the model handles low-concentration thresholds and high-concentration plateaus with equal precision.

\begin{figure}[htbp]
\centering
    \includegraphics[width=0.8\linewidth]{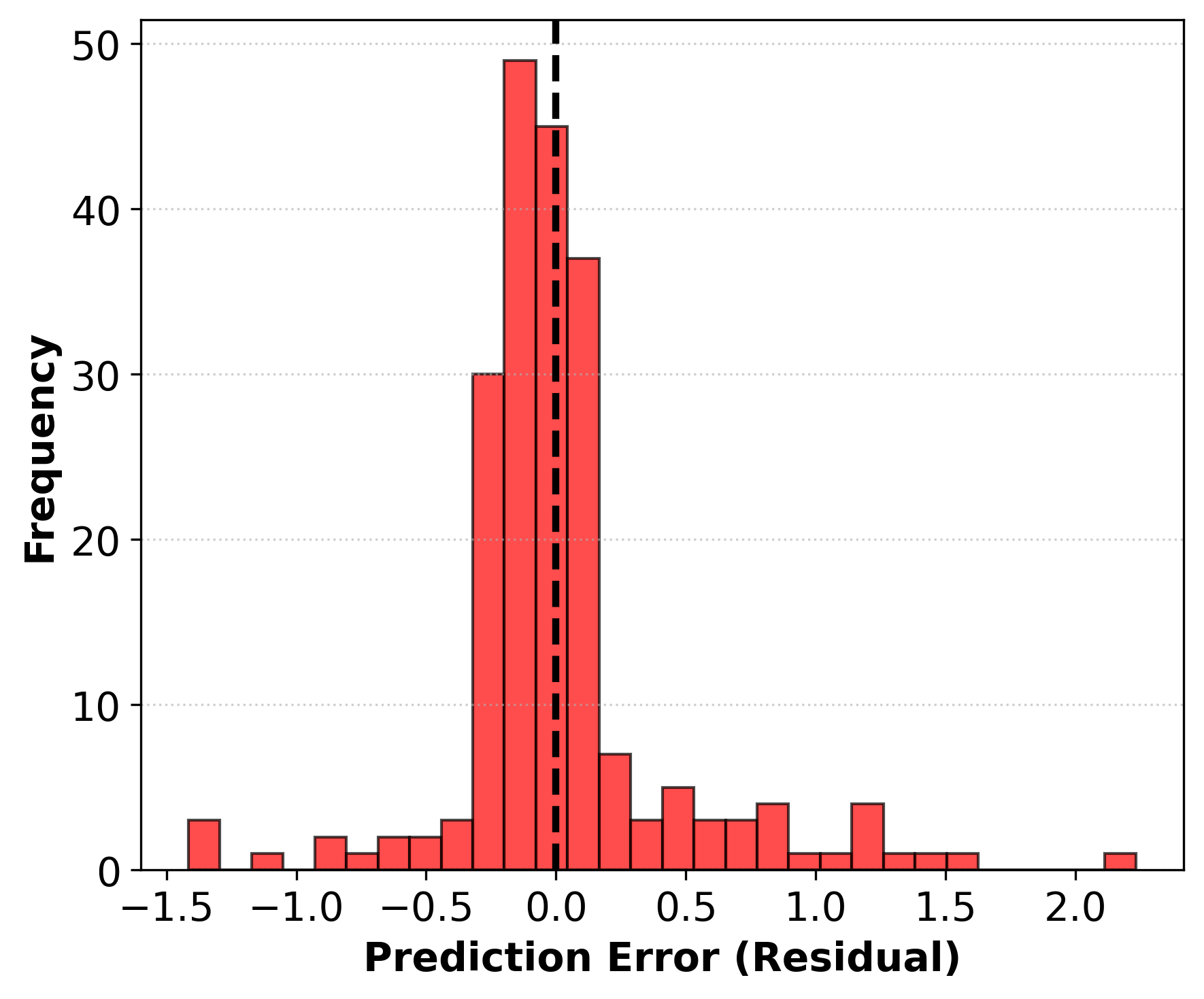}
    \caption{Residual error results for the VIANA model training.}
    \label{fig:VIANA_error}
\end{figure}

Ultimately, the residual error graph, Figure~\ref{fig:VIANA_error}, for VIANA shows a near-Gaussian distribution centered at zero, confirming that the synthesis of structural graph theory, character value transfer learning, and phenomenological kinetics provides a superior, robust, and efficient framework for assessing olfactory intensity. While the structural GCN suffers from ``information depletion'' when its character value input is compressed, the VIANA model benefits from ``signal distillation''.

To visualize these shifts in predictive accuracy, Tables~\ref{tab:gcn_comparison} and~\ref{tab:domain_comparison} highlight specific molecules from the test set across the different development phases.

\begin{table*}[htbp]
\centering
\caption{Comparison of GCN-based architectures on representative test molecules.}
\label{tab:gcn_comparison}
\begin{tabular}{lcccc}
\toprule
\textbf{Name} & \textbf{True Int.} & \textbf{Baseline} & \textbf{Char.\ value-enhanced} & \textbf{Reduced dim.} \\
\midrule
Octyl acetate              & 14.25 & 11.58 & 22.28 & 29.73 \\
Beta-pinene                & 16.56 & 10.49 & 7.61  & 8.19  \\
Dihydro-beta-ionone        & 1.66  & 10.85 & 8.92  & 3.81  \\
Sodium tetrahydrojasmonate & 14.69 & 11.28 & 18.61 & 10.19 \\
Coumarin                   & 16.29 & 11.51 & 14.88 & 9.71  \\
\bottomrule
\end{tabular}
\end{table*}

\begin{table*}[htbp]
\centering
\caption{Comparison of Domain-informed architectures leading to VIANA.} \label{tab:domain_comparison}
\begin{tabular}{lcccc}
\toprule
\textbf{Name} & \textbf{True Int.} & \textbf{Domain-informed} & \textbf{Char.\ value-enhanced} & \textbf{VIANA} \\
\midrule
p-Tolualdehyde & 15.19 & 14.99 & 15.85 & 15.10 \\
Stearyl alcohol & 1.30 & 1.17 & 1.06 & 1.34 \\
Isolongifolone & 1.36 & 1.22 & 1.28 & 1.49 \\
Alpha-Terpineol & 0.77 & 0.57 & 0.98 & 0.83 \\
beta-Santalol & 3.62 & 3.08 & 5.50 & 3.55 \\
3,5,5-Trimethylhexan-1-ol & 1.88 & 1.76 & 1.43 & 1.89 \\
\bottomrule
\end{tabular}
\end{table*}

The culmination of this research, embodied in the VIANA architecture, demonstrates that olfactory intensity can not be modeled or predicted as merely a function of molecular structure; rather, it requires a unified assessment that integrates molecular topology with character value and domain-informed physical constraints. By transitioning from the ``data-hungry'' and physically unconstrained GCN baseline to a domain-informed, distilled character value model, we have achieved a framework that captures the non-linear nuances of human perception. The high predictive fidelity ($R^2 = 0.996$) and the uniformity of error across all intensity ranges prove that the ``signal distillation'' provided by PCA is the key to unlocking the power of character value embeddings within a phenomenological context. VIANA does not just predict a number; it approximates the sigmoidal reality of human olfactory perception, providing a high-throughput tool that respects the physical ceiling of saturation and the sensitivity of detection thresholds.

However, it is important to notice that while VIANA represents a significant leap forward in olfactory modeling, its current validation is bounded by the existing experimental landscape. To further refine the model and expand its generalizability, a larger and more diverse repository of intensity data across a wider chemical space would be essential. Ultimately, the success of VIANA lays the groundwork for a new generation of ``chemically-aware'' and domain-informed AI models in sensory science.

\section{Conclusion}

This research addressed the challenge of predicting olfactory intensity by moving beyond traditional ``black-box'' approaches toward a biologically grounded synthesis of chemical and sensory data. The systematic progression through six modeling phases represented a deliberate journey of layering distinct forms of intelligence, structural, phenomenological, and character value, onto a neural network framework, effectively mirroring the biological hierarchy of human olfactory perception.

The journey began by establishing that structural knowledge via Graph Convolutional Networks (GCNs), while foundational, is insufficient for sensory tasks. Attempting to learn from a tabula rasa state, the structural baseline suffered from a ``data-starved'' condition, failing to differentiate molecules with similar frames but vastly different sensory intensities.

A pivotal breakthrough occurred with the introduction of phenomenological knowledge through Hill's Law. By transitioning to a domain-informed architecture, we directly encoded the fundamental physics of dose-response curves into the model's output head. The resulting surge in $R^2$ to 0.991 and the dramatic improvement in predictive fidelity demonstrated that a true assessment of intensity requires an inherent understanding of biological sigmoidal kinetics, specifically the interplay between detection thresholds and saturation points.

The integration of character value knowledge via Principal Odor Map (POM) embeddings added a qualitative ``vocabulary'' to the architecture. However, this phase revealed a critical architectural divergence: while semantic character enriched the structural model, it initially created ``information overload'' within the sensitive domain-informed head. This tension was resolved through the final innovation of optimized knowledge distillation. By applying Principal Component Analysis (PCA) to retain 95\% of the semantic variance, we filtered out latent noise and provided the model with a variance-optimized signal.

The culmination of this research is the VIANA architecture, which demonstrates that olfactory intensity is a complex intersection of molecular structure, odor character value, and phenomenological behavior. Through signal distillation, VIANA captures the non-linear nuances of human perception with unprecedented accuracy ($R^2 = 0.996$), providing a high-throughput tool that respects the physical ceiling of saturation and the sensitivity of biological detection thresholds.

While the success of the VIANA paradigm represents a significant leap forward in sensory AI, its current validation is bounded by the existing experimental landscape. To further expand its generalizability, a larger and more diverse repository of intensity data across a wider chemical space remains essential. Nevertheless, the integration of structural graph theory, distilled character value transfer learning, and phenomenological kinetics provides a superior framework that lays the groundwork for a new generation of chemically-aware and domain-informed olfactory models, effectively bridging the gap between molecular informatics and the human sensory experience.

\section*{Acknowledgments}

Climate Changes and Scent Heritage: The Urgent Need for Capturing and Preserving Olfactory Landscapes in a Changing World -- SCENTINEL, co-funded by the Joint Programming Initiative on Cultural Heritage and Global Change (JPI~CH). This work was financially supported by: FCT/MCTES (PIDDAC) UI/BD/154743/2023 (\url{https://doi.org/10.54499/UI/BD/154743/2023});
LSRE-LCM, UID/50020/2025 (DOI: 10.54499/UID/50020/2025);
and ALiCE, LA/P/0045/2020 (DOI: 10.54499/LA/P/0045/2020).

\section*{Author Contributions}
Conceptualization: L.P.Q.; B.M.A.M.; I.B.R.N.
Methodology: L.P.Q.; I.B.R.N.
Software: L.P.Q.
Validation: L.P.Q.
Formal analysis: L.P.Q.; I.S.C.B.
Resources: A.M.R.; I.B.R.N.
Data Curation: L.P.Q.
Writing -- Original Draft: L.P.Q.
Writing -- Review \& Editing: I.S.C.B.; A.M.R.; B.M.A.M.; I.B.R.N.
Visualization: L.P.Q.
Supervision: A.M.R.; B.M.A.M.; I.B.R.N.
Project administration: I.B.R.N.
Funding acquisition: A.M.R.; I.B.R.N.

\section*{Data Availability}
The data that support the findings of this study are available from the corresponding author (L.P.\ Queiroz) upon reasonable request.

\section*{Declaration of Competing Interest}
The authors declare that they have no known competing financial interests or personal relationships that could have appeared to influence the work reported in this paper.

\bibliography{cas-refs}

\end{document}